\documentclass[fleqn,usenatbib]{mnras}
\usepackage{newtxtext,newtxmath}

\usepackage[T1]{fontenc}

\DeclareRobustCommand{\VAN}[3]{#2}
\let\VANthebibliography\thebibliography
\def\thebibliography{\DeclareRobustCommand{\VAN}[3]{##3}\VANthebibliography}

\usepackage{graphicx}	
\usepackage{amsmath}	
\usepackage{newtxmath}
\usepackage{subcaption}
\usepackage{pdflscape}
\usepackage[dvipsnames]{xcolor}
\usepackage{booktabs}

\newcommand{\dsfr}{$\langle  SFR_{\textnormal{5}} \rangle \big/  \langle SFR_{\textnormal{200}} \rangle$}
\newcommand{\sfe}{SFE$_\text{H$_2$}$}

\title[The resolved recent SFH of NGC~628]{The recent star formation history of NGC~628 on resolved scales}

\author[M. Lomaeva et al.]{
Maria Lomaeva,$^{1}$\thanks{E-mail: maria.lomaeva.19@ucl.ac.uk}
Ilse De Looze,$^{1,2}$
Amélie Saintonge,$^{1}$
and Marjorie Decleir$^{3}$
\\
$^{1}$Dept. of Physics \& Astronomy, University College London, Gower Street, London WC1E 6BT, UK\\
$^{2}$Sterrenkundig Observatorium, Ghent University, Krijgslaan 281 - S9, 9000 Gent, Belgium\\
$^{3}$Space Telescope Science Institute, 3700 San Martin Drive, Baltimore, Maryland, 21218, USA
}

\date{Accepted XXX. Received YYY; in original form ZZZ}

\pubyear{2021}

\begin{document}
\label{firstpage}
\pagerange{\pageref{firstpage}--\pageref{lastpage}}
\maketitle

\begin{abstract}
Star formation histories (SFHs) are integral to our understanding of galaxy evolution. We can study recent SFHs by comparing the star formation rate (SFR) calculated using different tracers, as each probes a different timescale. We aim to calibrate a proxy for the present-day rate of change in SFR, dSFR/dt, which does not require full spectral energy distribution (SED) modeling and depends on as few observables as possible, to guarantee its broad applicability. To achieve this, we create a set of models in \texttt{CIGALE} and define a SFR change diagnostic as the ratio of the SFR averaged over the past 5 and 200 Myr, \dsfr, probed by the H$\alpha-$FUV colour. We apply \dsfr\ to the nearby spiral NGC~628 and find that its star formation activity has overall been declining in the recent past, with the spiral arms, however, maintaining a higher level of activity.
The impact of the spiral arm structure is observed to be stronger on \dsfr\ than on the star formation efficiency (\sfe). In addition, increasing disk pressure tends to increase recent star formation, and consequently \dsfr. We conclude that \dsfr\ is  sensitive to the molecular gas content, spiral arm structure, and disk pressure. The \dsfr\ indicator is general and can be used to reconstruct the recent SFH of any star-forming galaxy for which H$\alpha$, FUV, and either mid- or far-IR photometry is available, without the need of detailed modeling. 
\end{abstract}





\section{Introduction}\label{sect:intro}

Galaxies in the local Universe present a wide range of star formation (SF) activity. Galaxy-to-galaxy variations are illustrated by the bimodality of the population emerging as the blue and red sequence in the colour-magnitude distribution of galaxies. At the same time,  at fixed stellar mass, objects can be anything from quiescent early-type galaxies, to steadily star-forming late-types, all the way to starbursting systems \citep[e.g.][]{noeske_07, Gallazzi_07, Wuyts_11, Wetzel_12, van_der_Wel_14}. 
There are also significant variations in star formation activity within galaxies, with the star formation rate (SFR) and star formation efficiency (\sfe) observed to change with galactocentric radius \citep{Utomo_17, ellison_18}, spiral structure or presence of a significant stellar bulge \citep{leroy_13}, dynamically-induced features such as bars and disk asymmetries \citep{meidt_13}, or a combination of factors \citep{Belfiore_18, Medling_18}. 



This suggests that star formation activity is regulated by processes that operate on a variety of time and physical scales and that those processes depend on both local and global properties of the galaxies. Generally, local processes, such as stellar feedback, pressure, and turbulence, control the SFR \citep[e.g.,][]{sanchez_20}, while quenching (a significant or complete suppression of the SF) is driven by global ones, such as galaxy mass, environment \citep{Peng_10}, morphology \citep{Martig_09}, and Active Galactic Nucleus (AGN) feedback \citep[e.g.,][]{bluck_14, Bluck_22}. This complex interplay between various processes that affect the SF are imprinted in star formation histories (SFHs) of galaxies. Thus, in order to understand the evolution of galaxies we need to understand their SFHs.

Spectral energy distribution (SED) fitting represents one available method to estimate the SFH of a galaxy. However, it is challenging to obtain accurate results due to very diverse real SFHs that require large amounts of high-quality data and various assumptions to be pinpointed \citep[e.g.,][]{Papovich_01, Shapley_01, Muzzin_09, Conroy_13, Ciesla_16, Ciesla_17, Carnall_19, leja_19}, especially when short-term ($\sim$100~Myr) variations are concerned \citep[e.g.,][]{ocvirk_06, Gallazzi_09, Zibetti_09, leja_19}. To circumvent this issue in studies of recent SFH (< 1~Gyr timescales), one can compare relevant observations that probe short (5--10~Myr) and long (0.1--1~Gyr) time-scales in observed \citep{Sullivan_00, Wuyts_11, Weisz_12, guo_16, Emami_19, Faisst_19, wolf_19, wang_lilly_20, Byun_21, Karachentsev_21} and simulated \citep{sparre_17, Broussard_19, flores_velazques_21} galaxies.

A few such examples include \cite{Weisz_12} who measured H$\alpha$-to-FUV flux ratios for 185 nearby galaxies and found that more massive galaxies were best characterized by nearly constant SFHs, while low-mass systems experienced strong bursts lasting for tens of Myr with periods of $\sim$250~Myr. \cite{guo_16} studied SF through the ratio of H$\beta$-to-FUV-derived SFRs in 164 galaxies instead. They arrived at a similar conclusion that low-mass galaxies with $M_\star$ < 10$^9$~M$_\odot$ experienced a bursty SFH on a timescale of a few tens of Myr on galactic (global) scales, while galaxies with $M_\star$ > 10$^{10}$~M$_\odot$ formed their stars during a smooth continuous phase. \cite{Emami_19} investigated bursty SFHs in 185 local dwarf galaxies using H$\alpha$ and FUV observations and again found that the least massive galaxies (M$_\star$ < 10$^{7.5}$ M$_\odot$) in their sample experienced bursts with the largest amplitudes of $\sim$100 (the SFR at burst relative to the baseline SFR) and shortest duration (< 30~Myr). More massive galaxies with M$_\star$ > 10$^{8.5}$ M$_\odot$ experienced lower changes in SFR with amplitudes of $\sim$10 on >300~Myr time-scales. \cite{wolf_19} developed a quenching-and-bursting diagnostic using a combination of photometric colours  and measured the relative weight of A-type stars in a galaxy, probing quenching activity within $\sim$20~Myr. This tool allows to detect and reconstruct the SFH by detecting recent and local changes in the SFR. \cite{wang_lilly_20} studied the ratio of averaged SFRs between 5 and 800~Myr in MaNGA galaxies derived with H$\alpha$ emission, H$\delta$ absorption, and the 4000~\AA\,break. They found that the dispersion in this parameter, at a fixed galactic radius and stellar mass, is strongly anti-correlated with the gas depletion time. They also concluded that the scatter in SFR change parameter across the population is a direct measure of the temporal variability of the SFR within individual objects. \cite{Byun_21} used a similar tool, calculating the ratio between the SFR averaged over the past 10 and 100~Myr calculated via SED fitting, to investigate the observed H$\alpha$ flux deficit in the outer parts of two nearby, star-forming galaxies. They found that the drop in the flux ratio can be attributed to strong and short starbursts, followed by a rapid suppression of H$\alpha$ emission. \cite{Broussard_19} used simulations to define a burst indicator using SFRs on short ($\sim$10 Myr) and long ($\sim$100 Myr) timescales, which they suggested probing with H$\alpha$ and NUV emission. From the distribution of the burst indicator, they  concluded that its dispersion describes the burstiness of a galaxy population’s recent SF better than its average. The average should be close to zero, only to deviate from that if the galaxy population has an average SFH that undergoes a rapid enhancement or suppression. Finally, \cite{flores_velazques_21} studied burstiness through H$\alpha$ and FUV emission from the FIRE (Feedback in Realistic Environments) simulations. They confirmed that the SFRs are highly time variable for all high-redshift galaxies, while dwarf galaxies continue to be bursty to $z$ = 0. They also reaffirmed the use of the H$\alpha$-to-FUV ratio as an observational probe of SFR variability since they observed the SFR(H$\alpha$)/SFR(FUV) ratio decrease to < 1 when the true SFR decayed after a burst. 

\begin{figure}
	\includegraphics[width=\columnwidth]{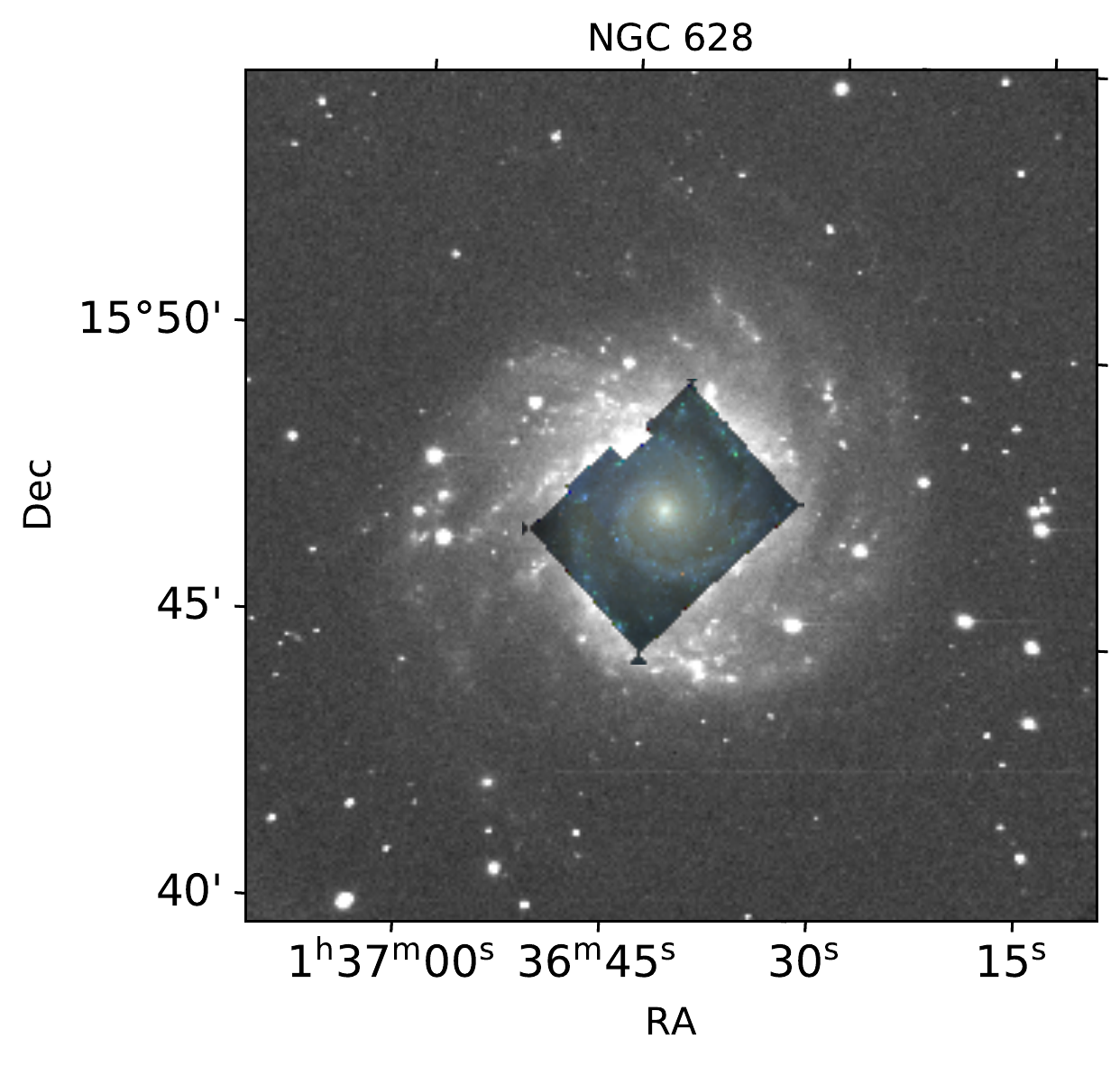}
    \caption{A \textit{V}-band image of NGC~628 overlaid with \textit{gri}-band photometric image from the PHANGS-MUSE sample \citep{Emsellem_22} to illustrate the region considered in this work. }
    \label{fig:ngc628}
\end{figure}

In this paper, we aim to investigate the recent SFH at low redshift and on resolved scales and calculate a SFR change diagnostic, \dsfr, which we define as the ratio between the SFR averaged over the past 5 and 200~Myr. To define our SFR change index, we generate a set of models and find an observable parameter, the H$\alpha-$FUV colour, with which we can probe \dsfr\,observationally. The main source of H$\alpha$ emission lines is ionised hydrogen gas in regions surrounding young stars, where electrons recombine with protons; thus, H$\alpha$ lines trace SFR on timescales of $\lesssim$5-10~Myr as the main source of ionising photons are massive OB-stars. UV emission of galaxies at wavelengths longwards of ~912~\AA~can directly trace the photospheric emission of young stars and reflect the SFR over the past few hundreds of Myr \citep{kennicutt_evans_12}. Since the relation between \dsfr\,and H$\alpha-$FUV is deduced theoretically, we can apply it to multiple star-forming galaxies to probe their resolved SFHs, specifically determining where within their discs star formation has been increasing or decreasing over the recent past. Our choice of 200~Myr as the reference timescale comes from the CIGALE models directly (see Section \ref{sect:cigale_results}), and is also supported by the findings in \cite{caplar_19} who showed that after 200~Myr SFHs of galaxies lose ‘memory’ of their previous SFH, that is, it becomes difficult to measure previous SFH through observational SFR diagnostics, such as H$\alpha$ and UV emission. 

To calibrate our SFR change diagnostic, we chose a galaxy with a large amount of observational data -- NGC~628, or M~74, shown in Figure \ref{fig:ngc628}. This galaxy is the largest in the NGC~628 group, where it is situated together with a peculiar spiral NGC~660 and their seven companions \citep{auld_06}. NGC~628 is a nearby grand-design spiral (SAc) galaxy seen almost perfectly face-on  \citep[inclination of 8.9$\degr$;][]{leroy_21_survey}. It lies at a distance of 9.59~Mpc, although its distance estimates vary between $\sim$7--10~Mpc \citep{kreckel_17}. NGC~628 has a global SFR of 1.74~M$_\odot$~yr$^{-1}$ \citep{leroy_21_survey}, stellar mass M$_\star$ = $2.2 \times 10^9$~M$_\odot$ \citep{leroy_21_survey}, and a weak metallicity gradient of $-0.0412\pm 0.05$~dex~kpc~$^{-1}$ \citep{kreckel_19}. An ultraluminous X-ray source was observed in NGC~628, indicative of a black hole with a mass of $\sim$$2\times10^3$~M$_\odot$ \citep{Liu_05}, although it could potentially be a stellar mass compact object instead \citep{alston_21}. Thus, there is currently no confirmed  AGN in NGC~628. \cite{headlight_cloud} observed a very bright molecular cloud traced in CO which spatially coincides with an extremely bright \ion{H}{ii} region in NGC~628, dubbed the ``headlight" cloud. It has a mass of $1-2 \times 10^7$~M$_\odot$ and is irradiated by a young (2--4~Myr) stellar population with a mass of  $3 \times 10^5$~M$_\odot$. \cite{Ujjwal_22} who studied the same region arrived at an age estimate of 16~Myr. The feedback from these young massive stars is destroying the headlight cloud. \cite{headlight_cloud} argue that the high mass of the cloud may be related to its location at a spiral co-rotation radius, where it receives a steady gas inflow due to a reduced galactic shear. 

NGC~628 has been the subject for a multitude of studies on galaxy formation and evolution \citep[e.g.,][]{Natali_92,Cornett_94, Elmegreen_06, Zou_11, Gusev_14, Grasha_15, Abdullah_17, Luisi_18,Rousseau-Nepton_18, inoue_21}. Having been selected as one of the first observational targets for the \textit{James Webb Space Telescope} \citep[JWST;][]{jwst}\footnote{\url{https://mast.stsci.edu/portal/Mashup/Clients/Mast/Portal.html}}, it also has a range of multiwavelength data available, making NGC~628 an interesting object for testing our SFR change index, before applying the method to a larger galaxy sample in future work. The morphological type and disk size of NGC~628 furthermore resembles our own Milky Way, and thus allows us to study what physical processes regulate whether an increase or decrease in star formation activity in Milky Way-type galaxies.

This paper is structured as follows: in Section \ref{sect:data} we describe the observations used in the analysis; in Section \ref{sect:dsfr_cigale_calibr}, we define and calibrate our SFR change diagnostic, \dsfr; in Section \ref{sect:results} we present resolved \dsfr\,in NGC~628 and show how this metric and SFE relate to the molecular gas reservoir, spiral arm structure, and mid-plane pressure; finally, in Section \ref{sect:discussion} and \ref{sect:conclusions} we discuss and conclude the main findings.

\section{Observations and data processing}\label{sect:data}

\subsection{Photometric observations}

We used multiwavelength photometric images ranging from the far-ultraviolet (FUV) to far-infrared (FIR). Those have been previously used in \cite{decleir_19} and include images obtained by the GALaxy Evolution eXplorer \citep[GALEX;][]{galex, morrissey_07}, the Sloan Digital Sky Survey \citep[SDSS;][]{sdss, eisenstein_11}, the InfraRed Array Camera \citep[IRAC;][]{irac} as well as the Multiband Imager \citep[MIPS;][]{mips} on-board \textit{Spitzer}  \citep{spitzer}, and the Photodetector Array Camera and Spectrometer \citep[PACS;][]{pacs} on-board \textit{Herschel} \citep{hershel}.

\cite{decleir_19} used the MIPS 24 $\micron$ image from the Spitzer Infrared Nearby Galaxies Survey \citep[SINGS;][]{sings}, while the remaining images were taken from the DustPedia Archive\footnote{\url{ http://dustpedia.astro.noa.gr}}. The DustPedia sample contains matched aperture photometric images of 875 nearby galaxies in over 40 bands \citep{davies17,clark18}. 

The SINGS data used in \cite{decleir_19} were reduced as in \cite{sings}, while the DustPedia images were reduced in a homogeneous manner, as described in \cite{clark18}. The image processing carried out by \cite{decleir_19} included subtraction of the background sky and foreground stars/objects, correction for the Milky Way extinction, convolution, rebinning, and uncertainty estimation. The images were convolved to the PACS~100~\micron\ resolution of about 7\arcsec (corresponds to a physical scale of 325~pc) and rebinned to a pixel grid of 7$\arcsec \times 7\arcsec$. This resolution allows to determine total infrared (TIR) emission, which we used for FUV attenuation corrections, while maintaining high resolution. \cite{decleir_19}  corrected the photometric images for the Milky Way (MW) extinction assuming a \cite{cardelli_89} curve and the Galactic extinction in the V band of $A_V= 0.188$ in NGC~628  (obtained from the IRSA Galactic Dust Reddening and Extinction Archive\footnote{\url{https://irsa.ipac.caltech.edu/applications/DUST/}}). 

The TIR luminosity density was calculated by \cite{decleir_19} following the formula from \cite{Galametz_13}:
\begin{equation}\label{eq:tir}
    S_{\textnormal{TIR}} = 2.162 \times S_{24} +0.185 \times S_{70} +1.319\times S_{100} \, ,
\end{equation}
with $S_{\textnormal{TIR}}$ the TIR luminosity density in units of W~kpc$^{-2}$, $S_{24}$, $S_{70}$, and $S_{100}$ the luminosity density in the MIPS~24~\micron, PACS~70~\micron, and PACS~100~\micron\, bands, respectively.


The UV radiation is sensitive to dust attenuation and should be corrected for such effects.  \cite{boquien_16} carried out a spatially resolved, multi-wavelength study of eight star-forming spiral galaxies from the KINGFISH survey \citep{kingfish}, including NGC~628. They relate the intrinsic UV luminosity emitted by the source with the observed one through the scaling coefficient $k$ and the observed IR luminosity in the corresponding band:
\begin{equation}
    L(UV)_\textnormal{int} = L(UV)_\textnormal{obs} + k \times L(IR)\,.
\end{equation}

By performing SED fitting, the authors deduced a relationship between the $k$ coefficient and several photometric colours to account for the variable impact of dust heated by old stellar populations (see their Table 4). We opted for:
\begin{equation}
    k = 0.943 -0.099 \times \textnormal{(FUV--IRAC 3.6)}\, ,
\end{equation}

setting $L(IR) = L(TIR)$ as one of the possible IR band options presented in \cite{boquien_16}. 
 
\cite{boquien_16} defined the TIR as the integral of dust emission over all wavelengths. With the TIR calibration in Equation \ref{eq:tir} from \cite{Galametz_13}, they can account for 99\% of the total variation of the resolved TIR brightnesses accounted for by their calibration. The modelling in \cite{boquien_16} was performed on scales of 0.5--1.7~kpc, and it is not recommended to apply this method to significantly higher resolution and finer spacial scales. This is generally compatible with our pixel size of 325~pc. For the recipe in \cite{boquien_16} to be valid, two additional constraints must be satisfied, that is 6.12 $\leq$ log $\Sigma$(TIR) $\leq$ 9.19 L$_\odot$~kpc$^{-2}$ and 0.44 < FUV--IRAC~3.6\micron\, < 5.98~mag, which is the case here.  

As a result, we estimated the average FUV attenuation to be $\langle A(FUV)\rangle = 1.60 \pm 0.20$~mag which was calculated from:
\begin{equation}
    L(FUV)_{\textnormal{corr}} = L(FUV)_{\textnormal{obs}} \times 10^{0.4 A(FUV)} \, ,
\end{equation}
where $L(FUV)_{\textnormal{corr}}$ and $L(FUV)_{\textnormal{obs}}$ are the corrected and observed FUV luminosities.

\subsection{Optical spectroscopy}


We used the optical spectra obtained with the Multi-Unit Spectroscopic Explorer \citep[MUSE;][]{muse}, which is an integral field spectrograph installed at the Very Large Telescope (VLT). These observations were available through the ESO Phase 3 Data Release\footnote{\url{http://archive.eso.org/scienceportal/home}}.

MUSE offers  a 1\arcmin $\times$ 1\arcmin \, field of view with a 0.2\arcsec\, pixel size. In total, 12 data cubes (IDs 094.C-0623(A), 095.C-0473(A), 098.C-0484(A)) with a spacial resolution of 0.7\arcsec--1.5\arcsec\, were analysed. The observations were made for a wavelength range between 4750--9350~\AA~and integrated over $\sim$40--46 min 
\citep{kreckel_16, kreckel_18}. The data reduction of the archival data was carried out using the MUSE pipeline  \citep[version 1.4 or higher][]{weilbacher_12, weilbacher_14, weilbacher_16}. 

We identified the foreground stars using the SIMBAD database \citep{simbad}; we then fitted a 2D Moffat profile to each star with the \texttt{mpdaf} package \citep{mpdaf1,mpdaf2} in \texttt{Python}. 
The estimated FWHM was used to define the size of the circular annulus and aperture centered at each star, which was implemented with the \texttt{Python} package \texttt{photutils} \citep{photutils}. We then masked out each star inside the \texttt{photutils} aperture using sigma clipping on the pixels inside the annulus. We performed the sigma clipping procedure using \texttt{astropy} \citep{astropy1, astropy2} with a 2$\sigma$ threshold. Once the pixels associated with the annulus that were lying outside of the 2$\sigma$ threshold were removed, we estimated the new mean and standard deviation; those we used to generate a set of random numbers drawn from a normal distribution to replace the star. This procedure ensured that the masks contained values that are related to the immediate surroundings of the stars. 

To correct for astrometric offsets between the cubes, we used an E-band (red) Digitized Sky Survey 1 (DSS 1) image. The shifts did not exceed 2\arcsec, in agreement with \cite{kreckel_19}. The data cubes were then merged using \texttt{montage}\footnote{\url{http://montage.ipac.caltech.edu}} and convolved with \texttt{mpdaf} in \texttt{Python} to match the PACS 100 \micron\, resolution of 7\arcsec\, assuming a Gaussian PSF. We also rebinned the MUSE data to a pixel grid of 7$\arcsec \times 7\arcsec$ to match the photometric images from \cite{decleir_19}. 

We then used the processed MUSE cubes to produce H$\alpha$ and H$\beta$ line emission maps. The fitting of each individual spectrum was done with \texttt{pPXF} \citep{ppxf}. From the H$\alpha$ line flux, we were able to first calculate the observed H$\alpha$ line luminosity, $L(H\alpha)_{\textnormal{obs}}$ and then translate it into the SFR, assuming a \cite{kroupa_01} IMF, as described in \cite{calzetti_summer}:
\begin{equation} \label{eq:sfr}
   \textnormal{SFR [M$_\odot$ yr$^{-1}$]}  = 5.5 \times 10^{-42} \times L(H\alpha)_{\textnormal{corr}} \, , 
\end{equation}
where $L(H\alpha)_{\textnormal{corr}}$ is the attenuation corrected H$\alpha$ line luminosity.

The H$\alpha$ dust attenuation correction factor, $A(H\alpha$), is calculated for each spaxel from the Balmer decrement, F(H$\alpha$)/F(H$\beta$). Assuming Case B recombination \citep{osterbrock_06}, we derive an expression for $A(H\alpha$):
\begin{equation}
 \begin{split}
    A(H\alpha) \, \textnormal{[mag]} = \frac{E(H\beta - H\alpha)}{k(H\beta) - k(H\alpha)} \cdot k(H\alpha) = \\ = \frac{2.5\log_{10} \left( \frac{1}{2.86} \cdot \frac{F(H\alpha)}{F(H\beta)} \right)}{\frac{k(H\beta)}{k(H\alpha)}-1} \, ,
 \end{split}
\end{equation}
where $\frac{k(H\beta)}{k(H\alpha)}$  is the reddening curve ratio of 1.53 for the \cite{calzetti_00} curve assuming a typical Milky Way value for the ratio of the total to selective extinction, $R_\text{V}$ = 3.1 \citep{cardelli_89}. This yielded an  average $\langle A(H\alpha)\rangle = 0.62 \pm 0.07$~mag.

The attenuation-corrected luminosity is obtained from the observed $L(H\alpha)_{\textnormal{obs}}$ as:
\begin{equation}
    L(H\alpha)_{\textnormal{corr}} = L(H\alpha)_{\textnormal{obs}} \times 10^{0.4 A(H\alpha)} \, .
\end{equation}

In addition to the internal dust attenuation, we corrected the H$\alpha$ and H$\beta$ maps for the MW dust extinction assuming a \cite{cardelli_89} curve with $R_\text{V}$ = 3.1. We note that we did not correct the H$\beta$ map for internal dust attenuation because it was only used to calculate the Balmer decrement.


We emphasise that we did not correct our observations for the H$\alpha$ emission from the diffuse ionised gas (DIG) which becomes important on resolved scales. In NGC~628, the DIG component might contribute $\sim$20--50\% of the H$\alpha$ emission \citep{kreckel_16, kumari_20}. The DIG emission is not directly connected to the recent SFR as it is produced through ionisation by young, massive stars ``leaking'' photons from \ion{H}{II} regions into the ISM, ionisation from old, post asymptotic giant branch stars, and shocks or hot-cold gas interface (see \citealt{manucci_21} and references therein). Thus, not all of the measured  H$\alpha$ emission in this work originates from the actual SF, and we might be overestimating the SFR, especially in the faint regions between the spiral arms.


\subsection{Molecular hydrogen observations}
We obtained $^{12}$CO(J=2$\rightarrow$1) line emission, hereafter CO(2–1), in NGC~628 from  the Physics at High Angular resolution in Nearby Galaxies (PHANGS) project\footnote{\url{https://sites.google.com/view/phangs/home/data}} \citep[PI: E. Schinnerer;][]{leroy_21_pipeline, leroy_21_survey}. To match the spacial resolution of the PACS~100\micron\, data, we used the line-integrated CO(2-1) intensity observations (moment-0 map) obtained with a  broad mask at 7.5\arcsec\, resolution and a 1$\sigma$ sensitivity of 5.5~mJy~beam$^{-1}$ per 2.54~km~s$^{-1}$ channel. 

The line-integrated CO(2-1) intensity,  $I_{\textnormal{CO}(2-1)}$, in K~km~s$^{-1}$ is converted into the molecular gas mass surface density, $\Sigma M(H_2)$, as:
\begin{equation}
   \Sigma M(H_2)\, \textnormal{[M$_\odot$ pc$^{-2}$]} = \alpha^{1-0}_{\textnormal{CO}} \cdot R^{-1}_{21} \cdot I_{\textnormal{CO}(2-1)} \cdot \cos i \, ,
\end{equation}
where $\alpha^{1-0}_{\textnormal{CO}}$ is the CO(1-0) conversion factor in M$_\odot$ pc$^{-2}$ (K km s$^{-1})^{-1}$, $R_{21}$ is the CO(2-1)-to-CO(1-0) line ratio, and $i$ is the inclination. We assumed $R_{21}$=0.61, which is the luminosity-weighted mean derived for NGC~628 in \cite{den_brok_21}, and adopt a constant Galactic value for $\alpha^{1-0}_{\textnormal{CO}}$ of 4.35 M$_\odot$ pc$^{-2}$ (K km s$^{-1})^{-1}$, as in \cite{bolatto_13}, since in NGC~628 almost all high-confidence $\alpha^{1-0}_{\textnormal{CO}}$ measurements are contained within a factor of two of the MW value of 4.4 M$_\odot$~pc$^{-2}$~(K~km~s$^{-1}$)$^{-1}$ \citep{sandstrom_13}.


\section{Defining the SFR change diagnostic} \label{sect:dsfr_cigale_calibr}

Our aim is to calibrate a prescription to estimate the rate of change in the SFR at the present time, dSFR/dt, based on simple observables rather than full SED modeling. Given both the stringent data requirements and important computational costs associated with the full modeling of SFH, a simple calibrated method relying on as few observables as possible has the advantage of a broader applicability. As discussed in Section \ref{sect:intro}, the H$\alpha$-to-FUV flux ratio is a commonly-used observable to infer the recent SFH of galaxies since H$\alpha$ probes the SFR on timescales of 5--10 Myr, while the FUV is sensitive to star formation on a timescale of $\sim$$100-300$ Myr. To identify possible degeneracies between H$\alpha-$FUV colour and different SFHs, as well as to calibrate the conversion between the observed colour and the rate of change in the SFR, we make use of synthetic models generated with \texttt{CIGALE}, as explained in Section \ref{sect:cigale_theory}. Our proposed SFR change diagnostic, \dsfr\, is calculated from the corrected H$\alpha-$FUV colour and calibrated to represent the ratio between the SFR averaged over the past 5 and 200~Myr (Section \ref{sect:cigale_results}).

\subsection{Combining H$\alpha$ line and UV continuum emission}\label{sect:ha_fuv_theory}

 To transform the H$\alpha$ line flux into flux density and relate it to the FUV photometry, we followed a procedure presented and used in \cite{boselli_16, boselli_18, boselli_21}. The procedure involves creating a pseudo-filter that relates the number of the Lyman continuum photons (LyC) that ionise \ion{H}{ii} regions to H$\alpha$ luminosity. \cite{boselli_16} did so by generating a grid of simulated galaxies and extracting their SED, measuring the flux density within the pseudo-filter, and comparing it to the number of ionising photons provided by the population synthesis models. This gave the expression:

\begin{equation}
    \textnormal{LyC [mJy]} = \frac{1.07\cdot10^{-37} \times L(\textnormal{H}\alpha)\,\textnormal{[erg s$^{-1}$]}}{(D\,\textnormal{[Mpc]})^2} \, .
\end{equation}

From this, the H$\alpha$--FUV colour can be derived as:
\begin{equation}
     \textnormal{H$\alpha$--FUV} = -2.5 \cdot \log_{10}(\textnormal{LyC [mJy]}) + 20 - \textnormal{FUV [mag]}\, .
\end{equation}
We used the attenuation-corrected H$\alpha$ and FUV observations in the calculations above. The average uncertainty of the corrected H$\alpha-$FUV colour is  0.06 mag.

\subsection{\texttt{CIGALE} models} \label{sect:cigale_theory}

\begin{figure}
    \centering
    	\includegraphics[width=\columnwidth]{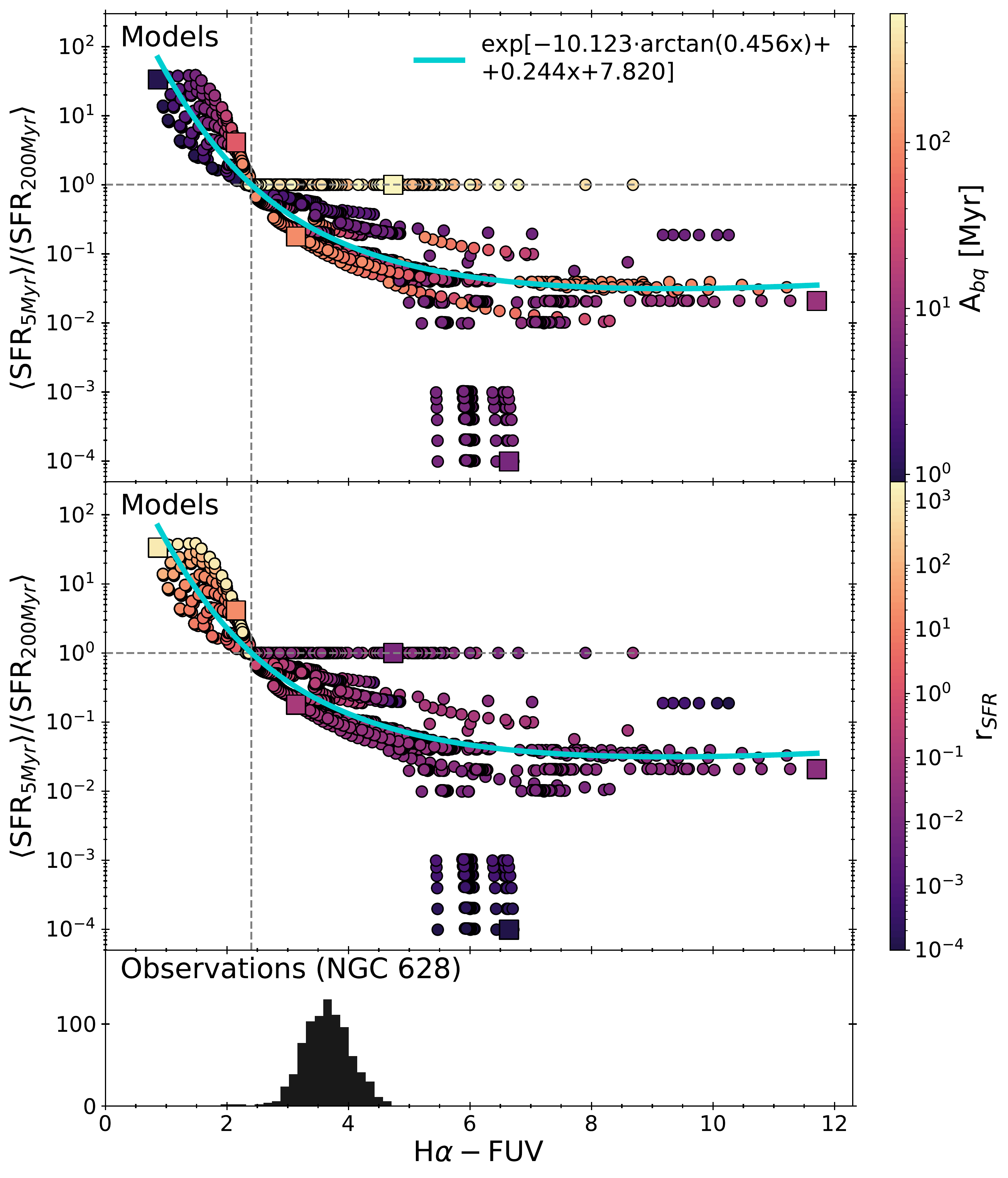}
        \caption{\textit{Top panel}: \texttt{CIGALE} models colour-coded according to the $A_\text{bq}$ parameter, i.e., how many Myr ago an increase or decrease in the SF took place. The dashed horizontal shows the \dsfr\, threshold between an enhancement and suppression of the SFR, while the dashed vertical line represents the same threshold at H$\alpha-$FUV = 2.4~mag. The turquoise line is the best fit function that describes the relation between \dsfr\, and H$\alpha-$FUV. The models with a high $A_\text{bq}$ along \dsfr\,= 1 and those creating a gap with \dsfr\,$\lesssim 2\times 10^{-3}$ were ignored during the fitting. The large squares indicate the models presented in Figure \ref{fig:cigale_sfhs}. \textit{Middle panel}: same as above but now colour-coded according to the $r_\text{sfr}$ parameter, which is the ratio between the SFR after and before a recent SF increase or decrease. Thus, $r_\text{sfr}>1$ indicates an enhancement in the SF, $r_\text{sfr}<1$ represents suppressed SF, and $r_\text{sfr}=1$ means a constant SFR. \textit{Bottom panel}: the distribution of the observed H$\alpha-$FUV colour in NGC~628 (corrected both for the internal dust attenuation and Milky Way extinction), which shows that NGC~628 has most recently undergone a suppression of the SF.}
        \label{fig:cigale_models_with_fit}
\end{figure}

To find a relationship between the observed H$\alpha-$FUV colour and the recent SFR change diagnostic, \dsfr, we generated a set of theoretical models in \texttt{CIGALE} \citep[Code Investigating GALaxy Emission;][]{cigale_1, cigale_2, cigale_3}. \texttt{CIGALE} models galactic spectra from FUV to radio wavelengths to estimate their physical properties, such as SFR, attenuation, dust luminosity, stellar mass, etc.

In \texttt{CIGALE}, we assumed a \cite{chabrier_03} IMF, while the population synthesis models for the stellar emission were taken from \cite{bruzual_03} (\texttt{bc03}). We selected a delayed SFH with a constant instantaneous increase or drop in the SFR  (\texttt{sfhdelayedbq}). The range of SFH parameters, such as the age, $A_\textnormal{main}$, and the e-folding time, $\tau_\textnormal{main}$, of the main stellar populations were taken from \cite{decleir_19}, who performed an SED fitting to individual pixels in NGC~628. The intensity of a SF increase or a drop in the SFR is parameterised by $r_\textnormal{SFR}$ which represents the ratio between the SFR after and before the  event. Thus, $r_\textnormal{SFR} < 1$ means a suppression in recent SF, $r_\textnormal{SFR} > 1$ means an increase in SF, and $r_\textnormal{SFR} = 1$ represents no change in the SFR. The $A_\textnormal{bq}$ parameter denotes how many Myr ago an enhancement or suppression in the SF occurred. The parameters of the model and the range of values explored are summarised in Table \ref{tab:cigale_param}.  We note that since we have corrected the H$\alpha-$FUV colour for attenuation effects, we can assume a dust-free environment for the models. 

\begin{table}
\caption{Parameter values used to generate \texttt{CIGALE} models. 
SFH parameters: $\tau_\textnormal{main}$, e-folding time of the main stellar population model in Myr; 
$A_\textnormal{main}$, age of the main stellar population in the galaxy in Myr;
$A_\textnormal{bq}$, how many Myr ago an increase or decrease in the SF occurred; 
$r_\textnormal{SFR}$, ratio of the SFR after/before an increase or decrease in the SF;
$SFR_\textnormal{A}$, value of SFR at t = 0 in M$_\odot$ yr$^{-1}$;
$norm$, flag to normalise the SFH to produce 1~M$_\odot$. 
Nebular emission parameters: 
$\log_{10} U$, ionisation parameter;
$f_\textnormal{esc}$, fraction of LyC photons escaping the galaxy;
$f_\textnormal{dust}$, fraction of LyC absorbed by dust; 
$W_\textnormal{lines}$, line width in km s$^{-1}$;
$emission$, flag to include nebular emission. 
Other parameters:
$IMF$, initial mass function;
$Z$, metallicity; 
$A_\textnormal{sep}$, separation age between the young and the old stellar populations in Myr;
$z$, redshift. 
}\label{tab:cigale_param}

\begin{tabular}{l c}
\hline 
\multicolumn{2}{l}{SFH parameters \texttt{[sfhdelayedbq]}} \\  
\hline
$\tau_\textnormal{main}$ & 1000, 2000, 4000, 5000, 6000, 7000, 8000  \\
$A_\textnormal{main}$  & 8000, 9000, 10000, 11000, 12000  \\
$A_\textnormal{bq}$ & \multicolumn{1}{p{4cm}}{1, 2, 3, 4, 5, 6, 8, 10, 15, 20, 30, 40, 50, 60, 70, 80, 90, 100, 200, 400, 600}  \\
$r_\textnormal{SFR}$  & \multicolumn{1}{p{4cm}}{0.0001, 0.0002, 0.0004, 0.0006,\newline 0.0008, 0.001, 0.01, 0.02, 0.04, 0.06, 0.08, 0.1, 0.2, 0.5, 1, 2, 5, 10, 20, 50, 100, 1000} \\
$SFR_\textnormal{A}$  & 1.0\\ 
$norm$  &  True \\
\hline

\multicolumn{2}{l}{Nebular emission \texttt{[nebular]}} \\ 
\hline
$\log_{10} U$ & --3.0 \\
$f_\textnormal{esc}$ & 0.0 \\
$f_\textnormal{dust}$ & 0.0 \\
$W_\textnormal{lines}$ & 300.0 \\
$emission$ & True \\
\hline

Other parameters \\ 
\hline
$IMF$ & 1 (Chabrier) \\
$Z$ & 0.02 \\
$A_\textnormal{sep}$ & 10 \\
$z$ & 0.00219 \\
\hline 
\end{tabular}
\end{table}


\subsection{Calibration of the SFR change diagnostic} \label{sect:cigale_results}

\begin{figure*}
\centering
	\includegraphics[width=\linewidth,height=\textheight,keepaspectratio]{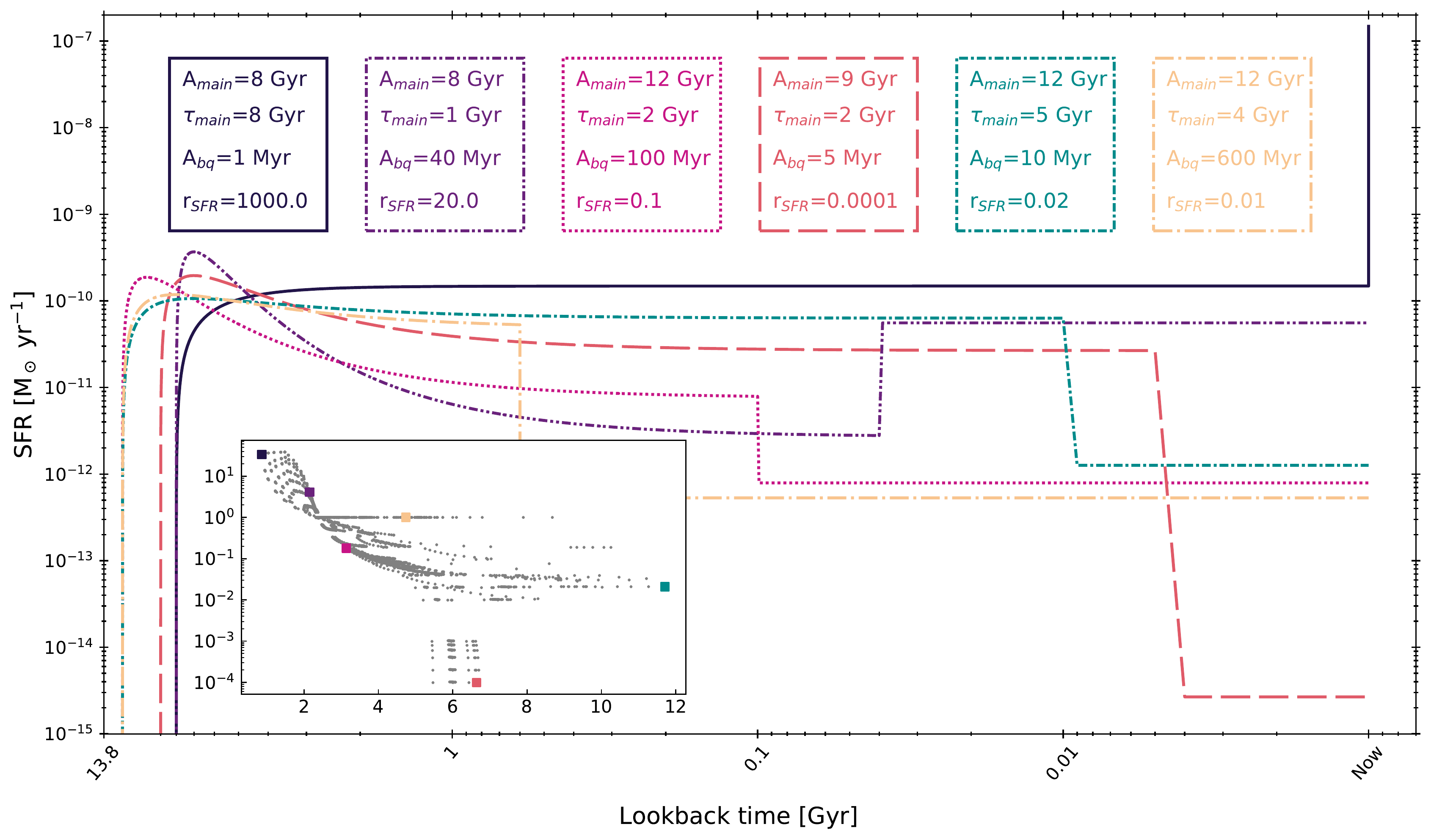}
    \caption{Selected SFHs of six \texttt{CIGALE} models with corresponding parameters. The inset plot shows the position of each model (coloured squares) in the \dsfr\, vs. H$\alpha-$FUV plane, as in Figure \ref{fig:cigale_models_with_fit}. The colour and line style of each box around the model parameters match those of the corresponding curve. In \texttt{CIGALE}, we followed the standard procedure and normalised SFHs such that the total stellar mass formed between the first and the last time step always equals to 1~M$_\odot$, hence, small SFR values.}
    \label{fig:cigale_sfhs}
\end{figure*}

In Figure \ref{fig:cigale_models_with_fit}, we show the models generated with \texttt{CIGALE} given the parameter values in Table \ref{tab:cigale_param}. Models that lie above \dsfr\,= 1 experienced an increase in the SF, while those below had a recent drop in the SFR. 

In total, we generated 16170 \texttt{CIGALE} models. In Figure \ref{fig:cigale_sfhs}, we plot a selection of modelled SFHs, their parameters, and their location in Figure \ref{fig:cigale_models_with_fit} (inset). These models were selected at random in different parts of the trend to illustrate how their SFHs compare w.r.t. each other. However, in Figure \ref{fig:cigale_models_with_fit}, we plotted 10627 models, excluding the models with $\tau_\textnormal{main}$ = 1~Gyr and $A_\textnormal{main}$ = 9--12~Gyr since those showed very little variation in H$\alpha-$FUV. That is rather expected because in these systems, the SFR peaked at very early times. In addition, the pixel-by-pixel SED fitting in \texttt{CIGALE} in \cite{decleir_19} did not result in any such SFHs with mostly old stars. Therefore, we are confident that we can exclude these model data points as they do not represent physical conditions encountered in nearby star-forming galaxies such as NGC~628.

 To obtain \dsfr, we extracted the SFR from the last 5 and 200~Myr (i.e., the last 5 and 200 steps in the model) and calculated the ratio of their averages. There is a certain dependence between the $r_\text{SFR}$ parameter and \dsfr\, diagnostic since $r_\text{SFR}$ is > 1 for an enhanced SF and < 1 for a suppressed SF, as seen in the middle panel of Figure \ref{fig:cigale_models_with_fit}. The advantage of the \dsfr\, diagnostic, however, is that we do not need to perform full spectral modelling, once the relationship between this parameter and the H$\alpha-$FUV colour is established. 

As shown in Figure \ref{fig:cigale_models_with_fit}, there are two other branches in addition to the main trend: (i) a cluster of points with a strongly suppressed \dsfr\, and $r_\textnormal{SFR}<10^{-3}$ and $A_\textnormal{bq}$~=~$5-6$~Myr at H$\alpha-$FUV $\approx$ 6; (ii) a branch with models that have $A_\textnormal{bq} \geq 100$~Myr and fall onto \dsfr= 1.

 The gap between the main population and branch (i) at H$\alpha-$FUV~$\approx$~6 arises due to a drop in the $\langle  SFR_{\textnormal{5 Myr}} \rangle$, which is potentially connected to the typical lifetime of H$\alpha$ ionising photons ($\sim$5~Myr), and their exhaustion in these models. This scenario is likely to be transient because of the very narrow range of $A_\textnormal{bq}$~=~$5-6$~Myr. In addition, the unrealistically drastic drop in the SFR, as shown by the long-dashed red line in Figure \ref{fig:cigale_sfhs}, also suggests that this scenario is unlikely to be observed in a real galaxy. Therefore, we do not expect pixels with H$\alpha-$FUV~$\approx$~6 to represent this particular case. 

Branch (ii), with \dsfr\,= 1, arises for models with $A_\textnormal{bq} \geq$ 100~Myr, meaning that in these models any changes in the SF activity occurred at relatively early times and the SFR has been constant over the past $>200$~Myr. This is also visible in  Figure \ref{fig:cigale_sfhs} (the yellow line). To break the degeneracy at fixed H$\alpha$-FUV colour between the models of branch (ii) and those on the main relation, we use additional photometric information to mask out pixels in the map of NGC~628 that are likely to be on the \dsfr\,= 1 branch. 
We opted for a double criterion based on the FUV--IRAC~3.6\micron\, colour and the equivalent width of the H$\alpha$ emission line, EW(H$\alpha$). Models with suppressed SF should be red in the  FUV--IRAC~3.6\micron\, colour, while having a low EW(H$\alpha$).
The final criteria were set to FUV--IRAC~3.6\micron\, > 3~mag and EW(H$\alpha$) < 5~\AA. 

In the next step, we isolated the main trend in the \dsfr\,vs. H$\alpha-$FUV plane and fitted a curve to it, as shown in Figure \ref{fig:cigale_models_with_fit}. The main trend contained 7786 models. The fit was performed using the \texttt{lmfit} package \citep{lmfit} in \texttt{Python}, assuming a natural logarithm of a model $F(x)$, that combines an arctangent and a linear function:
\begin{equation}
   \ln F(x) = a \cdot \arctan(bx) + (kx + m) \, ,
\end{equation}
where $a$, $b$, $k$, and $m$ are the fitted parameters. The fitting function is constrained to pass through the (2.4, 1) point since it is the transition between an enhancement and suppression in the SF activity, giving a relation:
\begin{equation}
    \begin{split}\label{eq:fit}
    \langle  SFR_{\textnormal{5}} \rangle \big/  \langle SFR_{\textnormal{200}} \rangle_\textnormal{fit} = \exp [ -10.123 \cdot \arctan (0.456 \cdot \\ \cdot (\textnormal{H}\alpha - \textnormal{FUV}))\, + 0.244 \cdot (\textnormal{H}\alpha - \textnormal{FUV}) + 7.820 ]\, .
    \end{split}
\end{equation}

\section{Results}\label{sect:results}


\subsection{Observed \dsfr\,in NGC~628}\label{sect:dsfr_results}

\begin{figure}
	\includegraphics[width=\columnwidth]{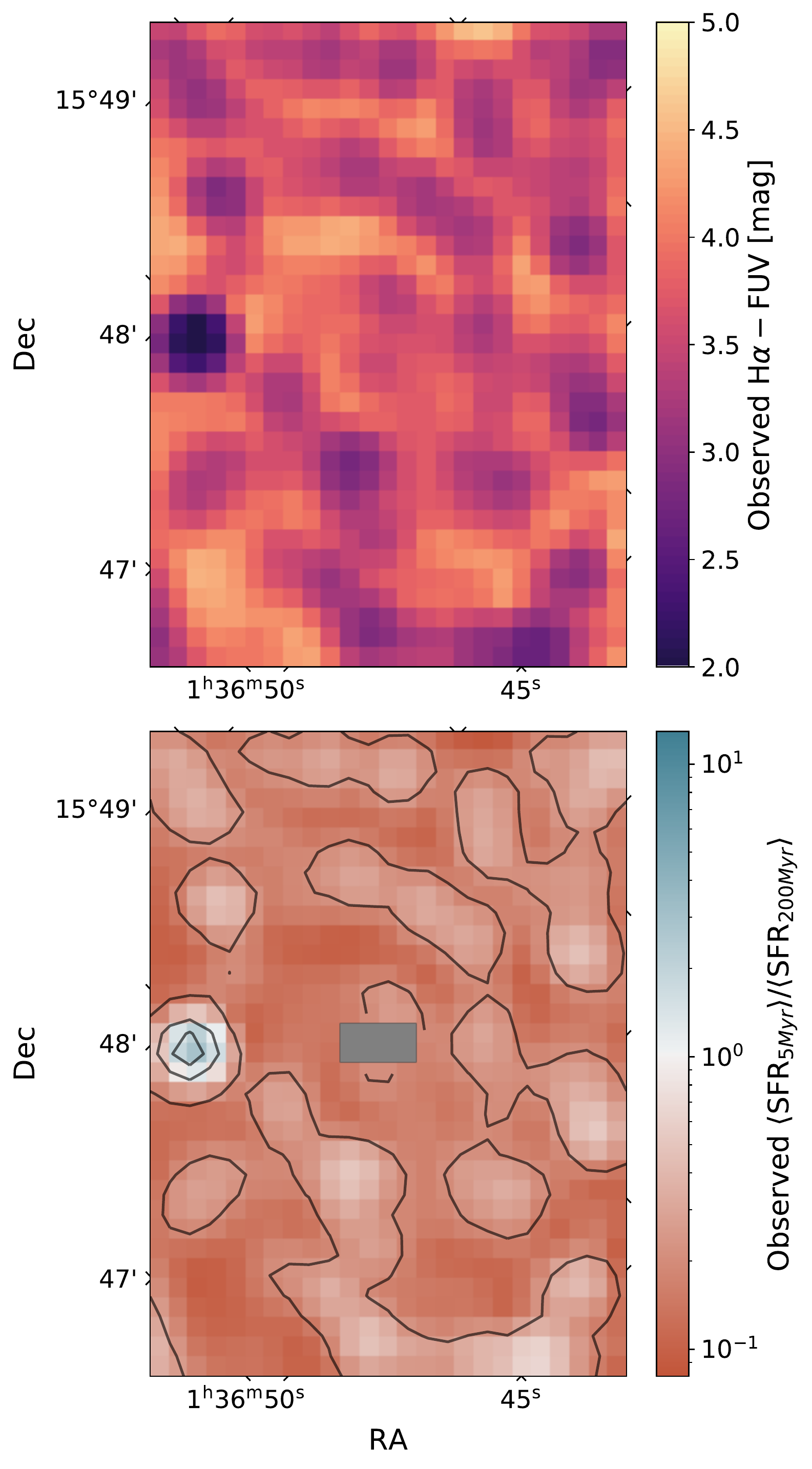}
    \caption{ \textit{Top panel}: H$\alpha-$FUV map corrected for internal attenuation and Milky Way extinction. \textit{Bottom panel}: observed \dsfr\,derived from the fitted relation shown in Figure \ref{fig:cigale_models_with_fit} and Equation \ref{eq:fit}. Blue colours show a recently increased SF, while red represents a recent suppression of the SFR. The contours follow  levels of \dsfr = 0.2, 1, 2, respectively. The central pixels were removed since they satisfy FUV--IRAC~3.6\micron\,> 3~mag and EW(H$\alpha$) < 5~\AA\, due to the absence of recent SFR changes.There are six pixels that pass through H$\alpha-$FUV = 2.4 within uncertainties, and thus, could correspond both to an event of an increase or decrease in the SF.}
    \label{fig:dsfr_dt_observed}
\end{figure}

In the top panel in Figure \ref{fig:dsfr_dt_observed},  we show the distribution of the observed H$\alpha-$FUV colour in NGC~628. The \dsfr\,parameter in this galaxy was derived by inserting the H$\alpha-$FUV colour corrected for the internal attenuation and MW extinction   into Equation \ref{eq:fit}. The corresponding \dsfr\,map is presented in the bottom panel of Figure \ref{fig:dsfr_dt_observed}. Pixels with blue colours are currently experiencing a phase where their SFR is increasing (\dsfr\,> 1), while those with red hues are currently experiencing a drop in the SF activity (\dsfr\,< 1).

The \dsfr\,map of NGC~628 shows that the galaxy is predominantly undergoing a phase of SFR suppression, although that decline is less rapid in the spiral arms. One region, east of the centre, stands out due to its strong recent burst -- that is the headlight cloud mentioned in Section \ref{sect:intro}. The masked pixels in the central region are those with FUV--IRAC~3.6\micron\, > 3~mag and EW(H$\alpha$) < 5~\AA\ where \dsfr\ is insensitive to the recent SFH, as revealed by the \texttt{CIGALE} models (see Section \ref{sect:cigale_results} and Figure \ref{fig:cigale_models_with_fit}).

\subsection{\dsfr\,and \sfe versus molecular gas reservoir}\label{sect:dsfr_sfe_gas}

\begin{figure*}
	\includegraphics[width=\textwidth]{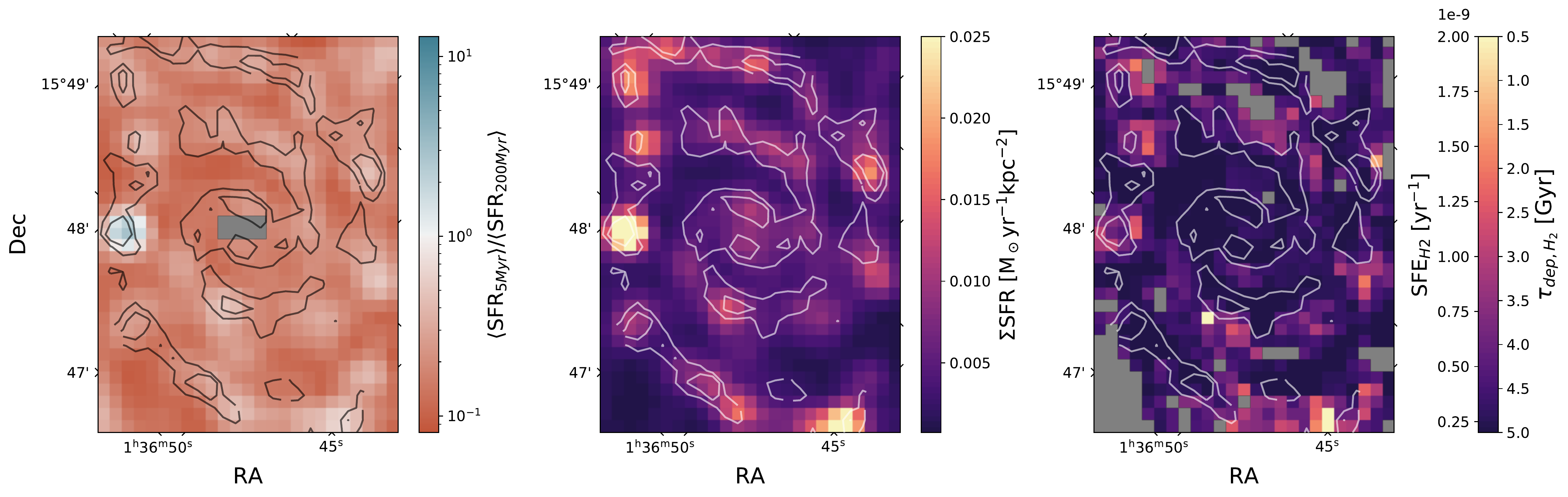}
    \caption{\textit{From left to right}: \dsfr, $\Sigma$SFR, and \sfe\,plotted with the contours representing $\Sigma$M(H$_2$) above the average noise level(4.5$\sigma$ and 8$\sigma$). The colour bar in the rightmost panel shows both the \sfe  values (left-hand side) in yr$^{-1}$ and the molecular depletion time, $\tau_\textnormal{dep}$, in Gyr (right-hand side). The masked pixels in the \sfe map are non-detections.}
    \label{fig:dsfr_dt_sfe_co_maps}
\end{figure*}

Knowing what areas have recently undergone a recent increase or suppression in the SFR, the next question we pose is how these areas are located with respect to the molecular gas reservoir. In Figure \ref{fig:dsfr_dt_sfe_co_maps}, we show \dsfr, SFR surface density, $\Sigma$SFR, and molecular SFE, \sfe = $\Sigma$SFR/$\Sigma$M(H$_2$), with the contours representing molecular gas mass surface density, $\Sigma$M(H$_2$). There is a strong overlap between $\Sigma$M(H$_2$) and both \dsfr\,(left panel) and $\Sigma$SFR (middle panel). There are however some areas with low-level star formation occurring outside of the regions of high molecular gas mass surface density, resulting in the relatively high star formation efficiency in the inter-arm regions (right panel). If one were to look only at the $\Sigma$SFR map, they might conclude that NGC~628 is forming new stars rather actively; while, \dsfr\,reveals that the SF has recently been decreasing in most regions. 

In the rightmost panel of Figure \ref{fig:dsfr_dt_sfe_co_maps}, we see that the high \sfe\ does not necessarily appear in the most gas-rich regions of the galaxy. For example, we observe a low level of the \sfe\ in the central region of the galaxy, despite a large H$_2$ gas reservoir. Looking at the inverse of the \sfe, the molecular depletion time $\tau_\textnormal{dep} \equiv~$\sfe$^{-1}$, we see an increase in its median value from 3.2~Gyr, averaged over the entire galaxy, to 5.2~Gyr in the bulge. Longer molecular depletion times in the inner part of NGC~628 have been observed before in \cite{kreckel_18}, for example. In our case, the decrease could be partly attributed to the radial variations of the $\alpha^{1-0}_{\textnormal{CO}}$ parameter, although its range is very limited in NGC~628 as it remains consistent with the MW value of 4.4 M$_\odot$~pc$^{-2}$~(K~km~s$^{-1}$)$^{-1}$ within a factor of two, as examined in \cite{sandstrom_13}. 

The innermost part of our Milky Way, the Central Molecular Zone (CMZ), contains $\sim$80\% of all dense molecular gas \citep{morris_96}. Yet, this region forms stars rather inefficiently \citep[e.g.,][]{longmore_13a}, in similarity with NGC~628, as the gas appears to be stabilised by turbulence \citep[][]{Krumholz_15}. \cite{orr_21} studied a set of Milky-Way mass spirals without AGN simulated with FIRE--2 \citep{Hopkins_18} and found that such a scenario could be a result of asymmetric and bursty galactic cores. \cite{moreno_21} used FIRE--2 interacting AGN-free galaxies and saw a fraction of their primary galaxies often experiencing low SFE levels, despite large boosts in cold-dense gas fuel. They explain it by the (stellar) feedback injecting turbulence into the ISM and preventing it from collapsing. 

Alternatively, the drop in the \sfe at the centre of NGC~628 could be due to stabilisation of the gas by the bulge, as proposed in \cite{Martig_09}. They propose that the shear, induced by the deeper gravitational well, prevents the gas from forming bound structures, suppressing star formation. \cite{davis22} indeed show that the molecular gas in the central region of galaxies is less fragmented in the presence of a massive bulge, resulting in significantly less star formation at fixed molecular gas surface density. 

\begin{figure}
	\includegraphics[width=\columnwidth]{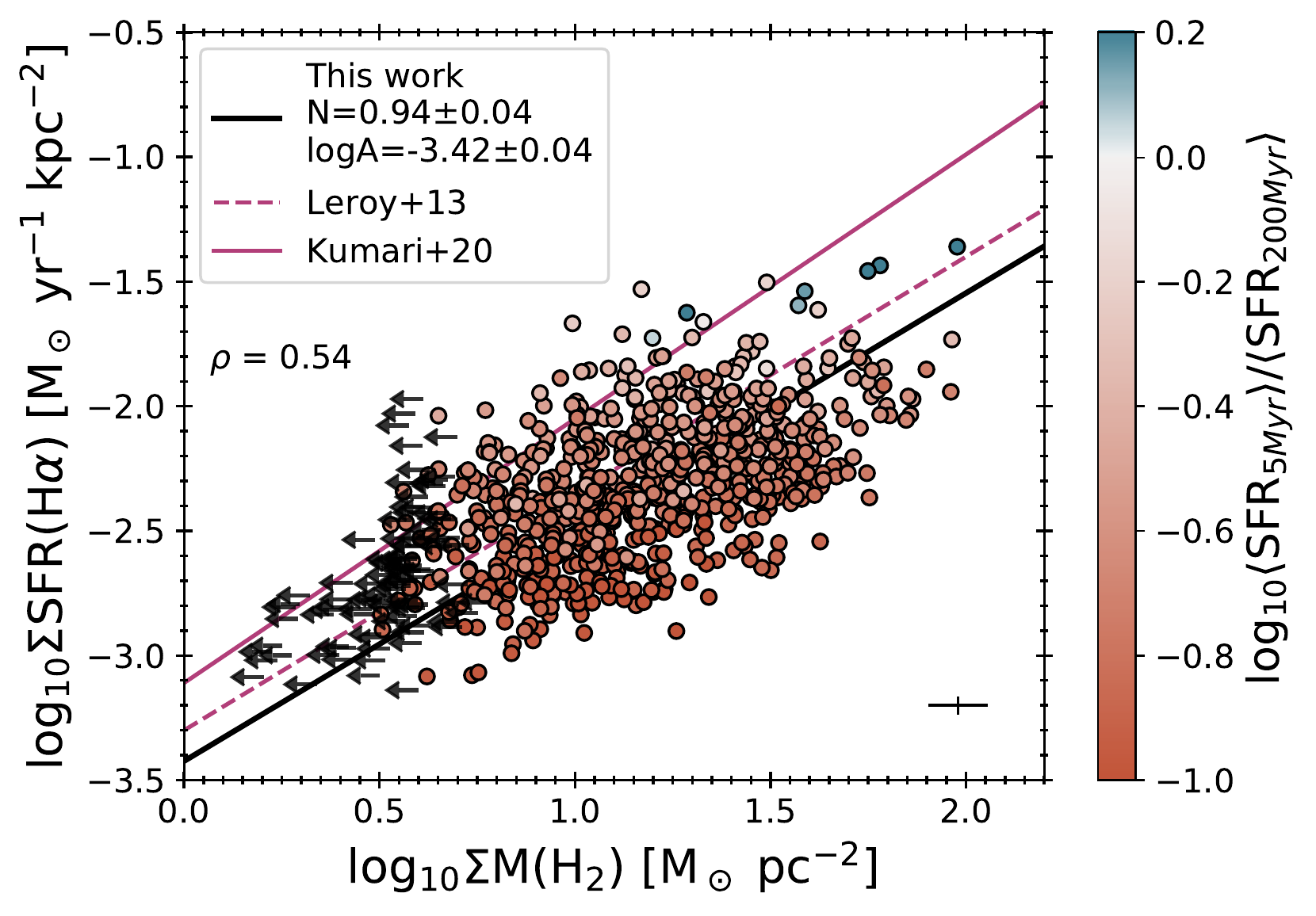}
    \caption{The resolved Kennicutt-Schmidt (K-S) relation with the arrows indicating 3$\sigma$ upper limits and a typical uncertainty in the lower right corner. $\rho$ denotes the Spearman correlation coefficient. The solid black line shows the fitted linear relationship, the dashed purple line indicates the K-S relation for NGC~628 from \protect\cite{leroy_13} with the slope of $N= 0.95 \pm 0.15$, while the solid purple line shows the corresponding results from \protect\cite{kumari_20} with $N= 1.06$. The non-detections were excluded from the calculations.}
    \label{fig:rKS}
\end{figure}

Another common approach for visualising the \sfe is by plotting the Kennicutt-Schmidt (K-S) relation, which we show in Figure \ref{fig:rKS}. We compare our results to \cite{leroy_13} as well as \cite{kumari_20} (not corrected for the diffuse ionised gas (DIG) emission). Both works studied a set of nearby galaxies on resolved scales, including NGC~628. Similarly as in \cite{kumari_20}, we create an unweighted linear fit for $\log_{10}(\Sigma \text{SFR}) = N \log_{10}(\Sigma\text{M(H$_2$})) + \log_{10}A$, where $N$ is the slope and $\log_{10}A$ is the intercept. The fit was performed using the orthogonal distance regression (ODR) algorithm in \texttt{Python}. This algorithm assumes normally distributed errors and finds the maximum likelihood estimators of parameters in measurement error models \citep{odr}. This gives a slope $N= 0.94 \pm 0.04$. In \cite{kumari_20}, the slope is $N= 1.06$ for NGC~628 and $N= 0.93 \pm 0.06$ for the average in their galaxy sample. \cite{leroy_13} fit a similar equation but using the Monte Carlo technique instead, accounting for uncertainties, upper limits, and intrinsic scatter. They get an average $N= 0.95 \pm 0.15$ for $\Sigma$SFR(H$\alpha$+24~\micron) and no cirrus subtraction. Thus, our slope value agrees with the literature rather well. 

The colour-coding according to the $\log_{10}$\dsfr\, in Figure \ref{fig:rKS} shows that at fixed $\Sigma$M(H$_2$), \dsfr\, increases with $\Sigma$SFR. In addition, the SFR enhancements reside at the highest $\Sigma$M(H$_2$)--$\Sigma$SFR end. For SFE, this is not always the case as illustrated by the headlight cloud in the rightmost panel of Figure \ref{fig:dsfr_dt_sfe_co_maps}, for example. There, the high SFR and molecular gas mass surface densities do not translate into a strong SFE. Moreover, SFE appears to be elevated in gas-poor and regions, where $\Sigma$SFR is suppressed.

The linear SFR relation, such as the one in Equation \ref{eq:sfr}, might break down at smaller scales, which would be directly translated into the scatter in the resolved K-S relation. Generally, \cite{kruijssen_14} formulated an uncertainty principle for the minimum scale size, for which the SFR relation still holds. In an idealised spiral galaxy, they found that such minimum scale is 500~pc. \cite{kreckel_18} showed that in NGC~628, the scatter in the depletion time at 300~pc scales is intermediate compared to the smallest (50~pc) and largest scales (2.4~kpc). The scatter we observe on 325~pc scales is $\sigma$($\log_{10}$(\sfe)) = 0.27~dex (cf. $\sigma$($\log_{10}$(\sfe)) = 0.33~dex at 300~pc scales in \cite{kreckel_18}). Our K-S relation  is also more strongly correlated than in \cite{kreckel_18} at 300~pc scales, with the Spearman correlation coefficient of 0.54 versus 0.25. This larger scatter could, however, originate from the DIG removal, which introduces additional complexity to the analysis.


In summary, elevated \dsfr\,values are generally observed in the areas with a large available molecular gas reservoir, which is expected since more SF should take place in the presence of cold gas, while the agreement between the \sfe\ and the molecular gas is not as clear. We also confirmed observationally that a higher \dsfr\ corresponds to an increase in $\Sigma$SFR at fixed $\Sigma$M(H$_2$), which is, again, expected as both $\Sigma$SFR and \dsfr\, were derived using H$\alpha$ emission. We interpret this as evidence that our \dsfr\,diagnostic is a more robust tool for predicting the recent SFH than the \sfe .

\subsection{\dsfr\,and \sfe in arms and interarms}
\begin{figure*}
	\includegraphics[width=\textwidth]{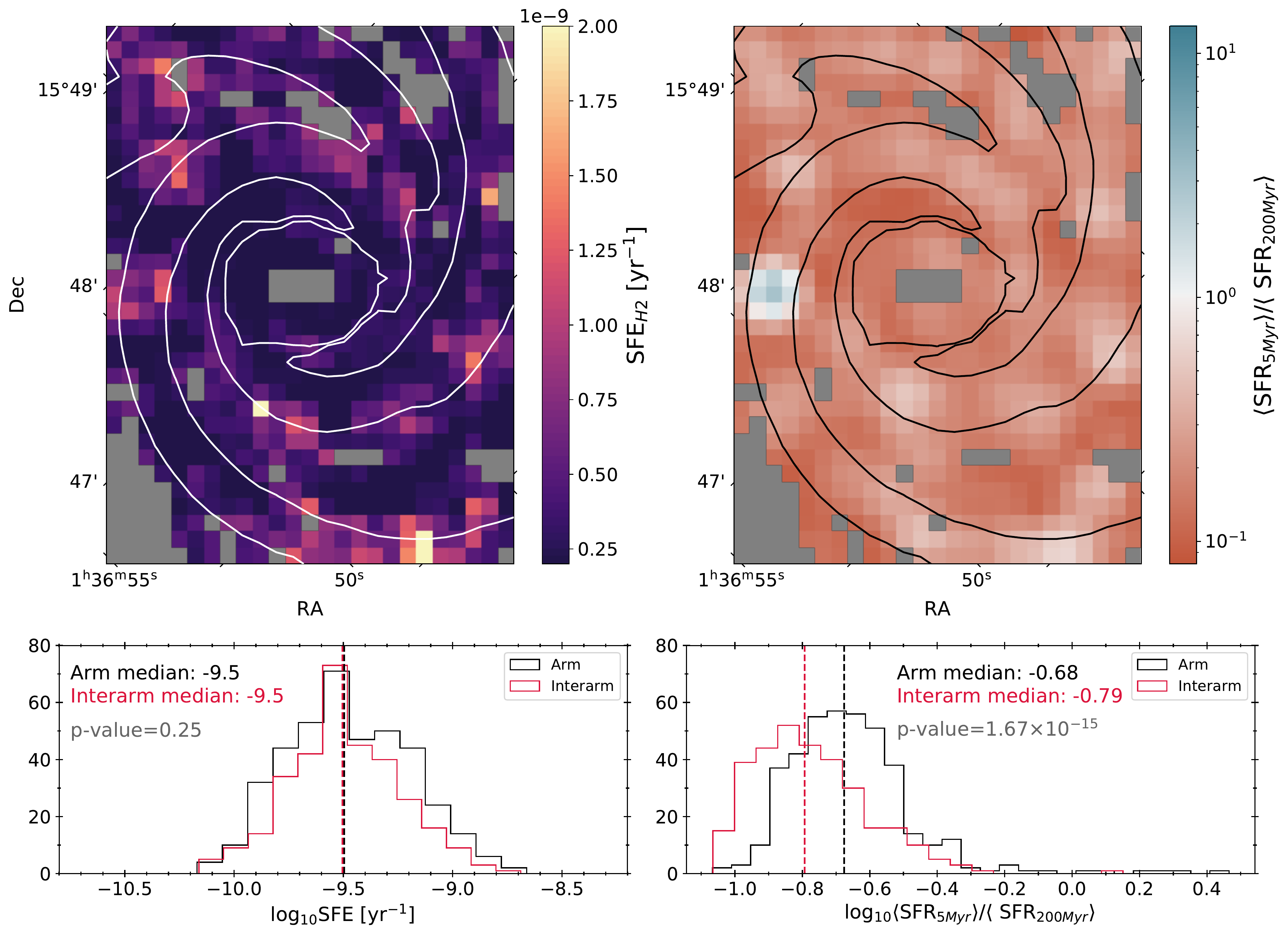}
    \caption{\textit{Top row}: the \sfe  and \dsfr\, maps with the spiral structure outline. \textit{Bottom row}: the arm and interarm distribution of the $\log_{10}$\sfe  and \dsfr\, shown with the p-values from the Kolmogorov-Smirnov test. The same pixels were masked in both maps (see Figure \ref{fig:dsfr_dt_observed} and \ref{fig:dsfr_dt_sfe_co_maps}) and were excluded from the statistical analysis. }
    \label{fig:arm_interarm}
\end{figure*}

In this section, we investigate the dependence between the spiral arm structure, \dsfr, and \sfe.  For this, we use a simple environmental mask for NGC~628 from \cite{Querejeta_21}\footnote{\url{http://dx.doi.org/10.11570/21.0024}} that was constructed from \textit{Spitzer} IRAC~3.6\micron\, images at $\sim$1.7\arcsec\, resolution. We convolved and rebinned the environmental mask to match our resolution and pixel scale. In the upper panels in Figure \ref{fig:arm_interarm}, we show the results for the \sfe and \dsfr\,maps separated into arm-interarm regions. \dsfr\,generally follows the spiral structure better than the \sfe, with high \sfe\ often occurring on both inner and outer edges of the spiral arms. 
This could be a result of the physical offset between the \ion{H}{ii} regions and molecular gas in NGC~628 which was studied by \cite{kreckel_18} who saw such offsets of $\ga$100~pc at a higher resolution ($\sim$1\arcsec\, both for H$\alpha$ and CO observations). \cite{egusa_09} found 33 offsets between the CO gas and H$\alpha$ emission in the inner part of the galaxy (7.2$\times$5.3\arcsec\, resolution for CO and 0.43\arcsec\, for H$\alpha$ data limited by 2\arcsec\, seeing). Given that a typical size of an \ion{H}{II} region is $\sim$35~pc \citep[e.g.,][]{kreckel_18}, we operate on physical pixel scales about ten times larger than that. This can average out the stochastic effects of the SF, and reduce the observed offset between the molecular gas and H$\alpha$ emission, but not entirely.


The lower panels in Figure \ref{fig:arm_interarm} show the distribution of \sfe\ and \dsfr\,in the arm-interarm regions. We applied the same masks to both maps (the $\Sigma$M(H$_2$) upper limits and the \dsfr\,masks discussed in Section \ref{sect:dsfr_results}) and performed the Kolmogorov-Smirnov test to investigate whether the arm-interarm populations are distinct for the two SF tracers. To do so, we used the \texttt{ks\_2samp} function in \texttt{Python}. The high p-value in the case of the \sfe\ suggests that these two populations are likely to originate from the same population. \dsfr, on the contrary, appears to be sensitive to the arm-interarm structure with a high statistical significance. Considering that the spiral arms reside in an increased gravitational potential, an increased presence of gas leads to a higher SFR in those regions. The \dsfr\, parameter, which is a ratio of SFR at two scales, is able to capture exactly that. 


\subsection{\dsfr\,and \sfe versus gas pressure} \label{sect:pressure_results}

Galaxy simulations show that $\Sigma$SFR is proportional to the ISM pressure in the disk \citep[see e.g.,][]{Gurvich_20}. In this section, we examine the dependence between \dsfr\,and \sfe\  and the pressure. We consider the expression for the mid-plane, or hydrostatic, pressure, following the procedure from \cite{elmegreen_89}:

\begin{equation}\label{eq:pressure}
     P_\textnormal{h} \approx \frac{\pi}{2} G \Sigma_{\textnormal{gas}}^2 + \frac{\pi}{2} G \frac{\sigma_\textnormal{g}}{\sigma_{\star,z}}\Sigma_{\textnormal{gas}}\Sigma_{\star} \, ,
\end{equation}
where $G$ = 4.301$\cdot$10$^{-3}$~pc~M$_\odot ^{-1}$~km$^2$~s$^{-2}$ is the gravitational constant, $\Sigma_\textnormal{gas}$ is the total ISM mass surface density (\ion{H}{i} + H$_2$), $\Sigma_\star$ is the stellar mass surface density, $\sigma_\textnormal{g}$ and $\sigma_{\star,z}$ are the total gas and vertical component of the stellar velocity dispersion, respectively. We assume $\sigma_\textnormal{gas}$ =  11~km~s$^{-1}$ when calculating the pressure to match the findings in \cite{leroy_08}. 

The first part of Equation \ref{eq:pressure} describes the self-gravity of the gas, while the second part represents the gas weight in the stellar potential well. This expression reflects the average behaviour of the midplane pressure needed to support the gas disk (ISM) from the gravitational collapse. We emphasise that we use $\Sigma$M(\ion{H}{i}+H$_2$), rather than $\Sigma$M(H$_2$), when calculating $P_\text{h}$. This is due to the fact that the gas in both phases contributes to the total pressure in the galactic disk. 

To estimate the vertical component of the stellar velocity dispersion, $\sigma_{\star,z}$, we follow the procedure in \cite{leroy_08}. It is based on the following assumptions: (i) the exponential stellar scale height, $h_\star$, does not vary with radius; (ii) $h_\star$ is related to the stellar scale length, $l_\star$, by  $l_\star$/$h_\star$= 7.3 $\pm$ 2.2; (iii) the galactic disk is isothermal in the z-direction. The above gives:

\begin{equation}
    \sigma_{\star,z} = \sqrt{\frac{2\pi G l_\star \Sigma_\star} {7.3}} \, ,
\end{equation}
where $l_\star$ = 2.3~kpc for NGC~628 \citep{leroy_08}. 

Since IRAC~3.6\micron\, observations are dominated by the emission from old stellar photospheres, we use the intensity in this band, $I_{3.6}$, to calculate $\Sigma_\star$, again following the recipe from \cite{leroy_08}:

\begin{equation}
    \Sigma_\star\, \textnormal{[M$_\odot$ pc$^2$]}= \Upsilon_\star ^K \bigg \langle \frac{I_K}{I_{3.6}} \bigg \rangle \cos i I_{3.6} = 280\cos i I_{3.6} \, ,
\end{equation}
where $\Upsilon_\star ^K$ = 0.5$\frac{M_\odot}{L_{\odot,K}}$ is the K-band mass-to-light ratio, I$_{3.6}$ = 0.55I$_K$, and I$_{3.6}$ is in MJy~sr$^{-1}$,  assuming a \cite{kroupa_01} IMF.

\begin{figure*}
	\includegraphics[width=\textwidth]{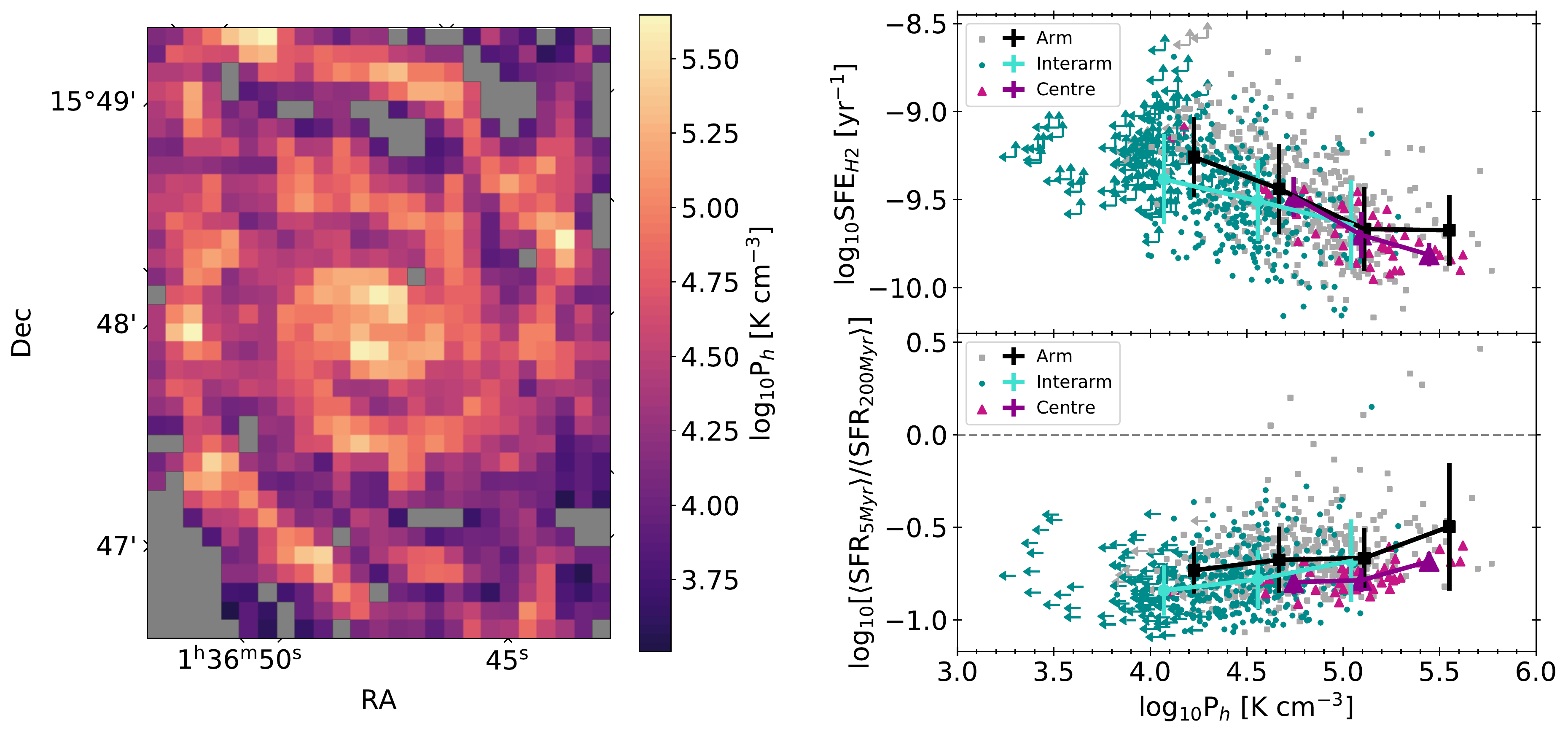}
    \caption{\textit{Left panel}: The distribution of the logarithm of the hydrostatic mid-plane pressure in NGC~628. The masked pixels are non-detections. \textit{Top and bottom right panels}: \sfe and $\log_{10}$\dsfr\,as a function of the hydrostatic mid-plane pressure, $\log_{10}P_\text{h}$ separated into the arm, interarm, and central pixels. The vertical solid lines represent the spread (one standard deviation) plotted at the median value within each bin. The 3$\sigma$ upper limits are shown as arrows and were not used for binning. The dashed grey line in the right panel marks the transition from an increased and suppressed star formation.}
    \label{fig:pressure}
\end{figure*}

We present the hydrostatic mid-plane pressure map of NGC~628 in the left panes of Figure \ref{fig:pressure}. Interestingly, in the top right panel of Figure \ref{fig:pressure}, the \sfe\ shows a decreasing trend with increasing pressure. The pressure, being proportional to the gas and stellar mass surface density, $\Sigma_\text{gas}$ and $\Sigma_\star$, respectively, increases towards the centre of NGC~628. As we previously noted in Section \ref{sect:dsfr_sfe_gas}, the \sfe\ in the centre is relatively low, resulting in the observed decreasing trend. Separating the observations into the arm, interarm and central regions using the simple environmental mask for NGC~628 from \cite{Querejeta_21}, this is exactly what we see: the central part of the galaxy stretches out to the high pressure end, while the \sfe\ remains suppressed. When we binned the data, the median values in each bin create otherwise similar decreasing trends in all three parts with large overlaps. 

Unlike the \sfe, the \dsfr\,diagnostic increases with increasing pressure, as shown in the bottom right panel of Figure \ref{fig:pressure}. This is in line with the expectations as higher pressure should increase recent SF activity, with the pixels corresponding to the strongest SF enhancements tending to lie at a higher pressure. We also see that the central part of NGC~628 has a systematically lower \dsfr\,than the arm-interarm regions.


As previously reported in \cite{kreckel_18}, we find that \sfe decreases towards the central, higher pressure region of NGC~628.  This could be due to the stabilising effect of the stellar bulge against fragmentation and collapse \citep[e.g.,][]{Martig_09} or (stellar) feedback effects \citep{moreno_21, orr_21}, discussed in Section \ref{sect:dsfr_sfe_gas}. On the other hand, we observe \dsfr\, to increase in the regions of highest pressure within the galactic disc.

\section{Discussion}\label{sect:discussion}


\subsection{The role of spiral arms in regulating SF}



The question of whether spiral arms can boost SF activity, and consequently SFE, has been debated over several decades. Early works suggested that the arms could boost the SFE by propagating a supersonic density wave w.r.t the ISM \citep[e.g.,][]{roberts_69, roberts_75, gittins_04}. For example, \cite{cepa_90} used H$\alpha$ emission and \ion{H}{I} surface density to derive SFE in NGC~628. They found enhancements in the arm-interarm ratios, suggesting that spiral arms can indeed boost the SFE. \cite{vogel_88} and \cite{lord_90} used CO emission and total gas surface density, respectively, combined with H$\alpha$ emission and found similar results for M51. \cite{knapen_96} looked at the \ion{H}{I}, CO, and H$\alpha$ emission distribution in the grand-design spiral NGC~4321, and determined that the total gas SFE is about three times higher compared to the interarm regions due to the compression of the gas by a density wave shock. \cite{seigar_02} used H$\alpha$ and K-band light in the arms for a sample of 20 spiral galaxies, again, confirming a SFE boost in the arms. Based on the idea that GMCs are formed by the spiral shock wave and therefore appear to be fixed to the spiral pattern \citep{egusa_04, egusa_09}, \cite{gao_21} argue that the compression of the molecular gas can even lead to differences in SFE between the front (leading) and back (trailing) part of a spiral arm, potentially breaking the K-S relation at higher resolutions. 

On the other hand, some observations have been in favour of the alternative theory, which states that any enhancement in the SFR along the spiral arms is simply due to a higher concentration of gas in these regions. This gas remains longer there which favours SF \citep[e.g.,][]{elmegreen_elmegreen_85, elmegreen_elmegreen_86}. Thus, the spiral arms, as such, do not enhance the SFE.  \cite{foyle_10} studied the SFR traced by a combination of FUV and 24\micron\, emission and the molecular \sfe in NGC~628, among other galaxies, at spacial scales of 250--600~pc. They found no evidence that spiral arms would lead to a higher \sfe through shocks. \cite{kreckel_16} examined 391 \ion{H}{ii} regions at 35~pc resolution in arm and interarm regions in NGC~628. They used optical MUSE observation to estimate the SFR and total gas mass through dust attenuation. The authors  found no difference between the SFE within the \ion{H}{ii} clouds in the arm and interarm environments. This conclusion is, however, very sensitive to the corrections preventing the contamination of the \ion{H}{ii} emission by DIG. \cite{schinnerer_17}, who studied SF in M51, found that it did not only occur inside the spiral arms, but also immediately outside of them, in so-called spurs, which could not be explained by shocks induced by a density wave. \cite{Querejeta_21} arrived at a similar conclusion after having studied a sample of 74 nearby PHANGS galaxies and compared the depletion time in the arm versus interarm available in 26 of those. In fact, they saw that in some cases the depletion time was longer in the spiral arms. This, again, indicates that the spiral arms accumulate gas and SF, but do not necessarily render the SF more efficient. 

There have been other attempts to explain the variations of the SFE along the galactic disk. For instance, \cite{meidt_13} examined the depletion time in M51 through the CO, H$\alpha$ and 24\micron\, emission and found a variation in $\tau_\textnormal{dep}$ along the galactic disk. They explain this finding with the changes in the gas streaming motions caused by gravitational deviations from axisymmetry in the disk. These deviations can lead to large streaming motions which can stabilise giant molecular clouds (GMCs) and prevent them from collapsing.

There is also a possibility that the spiral structure in NGC~628 does not (fully) originate from a stationary density wave. Instead, it could be a transient feature caused by swing amplifications, that is, local amplifications in a differentially rotating disk, \citep{Toomre_81}. A static wave should produce an age gradient across the spiral arms, with young stars trailing behind the spiral arm inside the co-rotation radius (CR), where the matter and the spiral structure have the same angular speed, and the in front of it, outside of the CR \citep[for more details see, e.g.,][]{Martinez_Garcia_09}. However, this has not been observed in NGC~628 \citep{Shabani_18, Ujjwal_22}.


In our case, we see that the \sfe varies along the spiral arms in NGC~628, as shown in the left panel in Figure \ref{fig:arm_interarm}, with some differences between the trailing and leading side of the spiral arms. This would suggest that the spiral structure does increase the \sfe. However, looking at the distributions of the \sfe in the arms and interarms in Figure \ref{fig:arm_interarm}, there is no significant statistical evidence that these regions are any different, which indicates that the spiral structure does not have an effect of the \sfe. \dsfr, on the other hand, is more sensitive to the spiral structure, again see Figure \ref{fig:arm_interarm}. This is expected since the density wave concentrates the gas along the spiral structure, thus, we can expect an increase in the recent SFR in the arms compared to the interarms. Therefore, our results are more in line with the previous SFE observations of NGC~628 from \cite{foyle_10} and \cite{kreckel_16}, for example. This means that the spiral arm structure increases the amount of gas and SFR but not necessarily the efficiency with which molecular gas is transformed into stars.

In the analysis, we did not remove the DIG emission. Its removal is method-dependent and would introduce additional uncertainties. Moreover, if we remove DIG, we should also remove diffuse CO gas which does not directly participate in SF either. That would add another level of complexity and enhance uncertainties even further. Moreover, the conclusion that we draw about the SFE and \dsfr\, in the arm-interarm environments would only be reinforced without the DIG contribution.  We likely overestimate the SFR in the interarm regions, and thus, the contrast between the spiral arms and interarm regions would be even stronger.

Considering the results on the mid-plane gas pressure in Section \ref{sect:pressure_results}, the \sfe in NGC~628 does not grow with higher pressure, potentially due to the stabilisation of the gas at the centre of the galaxy. \dsfr\,behaves more intuitively, as higher pressure leads to elevated \dsfr. An increase in pressure that can trigger SF may not be immediately captured by the \sfe\ measurement, if there is a large gas reservoir available in that particular part of the galaxy, or if the SFR indicator used represents the time average over a longer period. 

\subsection{Comparison with other SFR change diagnostics}
\begin{figure*}
	\includegraphics[width=\textwidth]{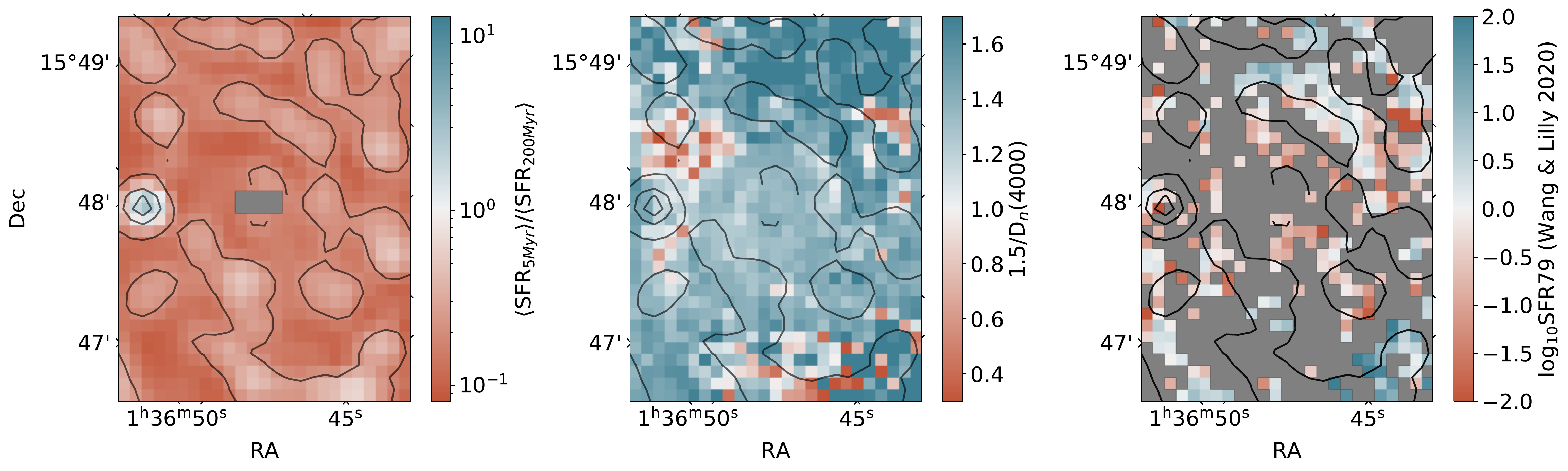}
    \caption{\textit{Left to right}: the comparison between \dsfr, D$_\text{n}(4000)$, and $\log_{10}$SFR79 \protect\citep{wang_lilly_20}, probing recent changes in the SF. The contours show elevated levels of \dsfr\,as in Figure \ref{fig:dsfr_dt_observed}. Recent SF enhancements are shown in blue, while the suppression is in red. The $D_{\text{n}}$(4000) parameter was defined as $1.5/D_{\text{n}}$(4000) following \protect\cite{kauffmann_03} to match the colour scale of the other diagnostics.  }\label{fig:dsfr_dt_sfr79_d4000}
\end{figure*}

In this section, we compare our SFR change index to other diagnostics that were similarly calibrated to identify recent changes in the SF activity. One such diagnostic was presented in \cite{wang_lilly_20}, and makes use of the H$\alpha$ equivalent width, EW(H$\alpha$); the Lick index of the H$\delta$ absorption, EW(H$\delta$)$_{\text{A}}$; and the 4000~\AA\, break, $D_{\text{n}}$(4000). The H$\alpha$ emission traces star formation on short timescales ($\sim$5\,Myr), while the H$\delta$ absorption is sensitive to SF during the last 1\,Gyr. The $D_{\text{n}}$(4000) break is sensitive to the average (luminosity-weighted) stellar age within 2\,Gyr. The combination of these three diagnostics has been calibrated in \cite{wang_lilly_20} to trace the SFR change parameter, $\langle SFR_{5\,\text{Myr}}\rangle / \langle SFR_{800\,\text{Myr}}\rangle$ or ``$\log_{10}$SFR79” following their notation. Their SFR change parameter is particularly sensitive to changes in the SF activity that have occurred during the last 5\,Myr, relative to the long-term SF activity on 800\,Myr timescales. Given that this SFR change parameter identifies changes in the recent SF activity with respect to the SF on much longer timescales ($\sim$800\,Myr), we may expect deviations from our \dsfr\,diagnostic. Nevertheless, it remains a good exercise to compare both SFR change parameters, and to study possible differences between both methods.

Since the MUSE spectra do not cover either the H$\delta$ line nor the 4000~\AA\, break, we use the PINGS IFS data cube from \cite{pings} to calculate the SFR change index presented in  \cite{wang_lilly_20}. We first regridded the interpolated 3D data cube from PINGS to the 7\arcsec\, pixels used in the analysis with the MUSE spectra to enable a one-to-one comparison of different regions.
We used \texttt{pPXF} to fit the stellar absorption and gas emission lines and inferred the H$\alpha$ (corrected for stellar absorption) and H$\delta_{\text{A}}$  (corrected for gas emission) equivalent widths (EWs) from these fits. We adopted the bandpasses as defined in \cite{balogh_99} to calculate $D_{\text{n}}$(4000) and EW(H$\delta$)$_{\text{A}}$. We used the calibration coefficients for solar metallicity presented in Table 1 of \cite{wang_lilly_20} to calculate the SFR change parameter $\log_{10}$SFR79. We only applied the calibration to the pixels in which both the H$\alpha$ and H$\delta$ line were detected at sufficient signal-to-noise, i.e., SNR $\ge3$.
 
Although the use of EWs in the $\log_{10}$SFR79 diagnostic limits the effects of dust attenuation, the differential attenuation between young ($<$10\,Myr) and old stellar populations — or between line and continuum emission — may bias the $\log_{10}$SFR79 diagnostic. We applied the calibrations for dust corrections to EW(H$\alpha$), EW(H$\delta$)$_{\text{A}}$, and $D_{\text{n}}$(4000) from \cite{wang_lilly_20} that were inferred based on the \cite{cardelli_89} dust curve and the assumption that stars younger than 10\,Myr are more heavily obscured (i.e., they experience roughly 3 times higher attenuation than old stars). These dust-correction calibrations require an estimate of the reddening experienced by young stars — $E(B-V)_{\text{young}}$ — which we derive from the Balmer decrement.

Figure \ref{fig:dsfr_dt_sfr79_d4000} shows the \dsfr\,(left), $D_{\text{n}}$(4000) (middle) and $\log_{10}$SFR79 (right) maps. 
In the middle panel, we actually show 1.5$/D_{\text{n}}$(4000), since \cite{kauffmann_03} first indicated that  $D_{\text{n}}$(4000) < 1.5 is characteristic for stellar populations younger than 1~Gyr. Thus, 1.5$/D_{\text{n}}$(4000) > 1 would imply an increase in SF and a suppressed SF otherwise. We do see that 1.5$/D_{\text{n}}$(4000) > 1 in most parts of the galaxy, indicating an increased SF. Our \dsfr\,index is on the contrary mostly decreasing, which can be explained by the different timescales probed by these metrics. Moreover, the 1.5$/D_{\text{n}}$(4000) metric is unable to detect the headlight cloud. This discrepancy could be eliminated if the SFR reached its peak between 5~Myr and $\sim$1~Gyr ago and its intensity in the headlight cloud was typical across almost the entire galactic disk. In fact, \cite{MacArthur_09} suggest that there was indeed a boost in SF $\sim$1~Gyr ago that is responsible for $\sim$40\% of the central stellar mass in NGC~628. 

The $\log_{10}$SFR79 index calculated as in \cite{wang_lilly_20} in the right panel shows a decreasing SF in the central part of the galaxy (i.e., $\log_{10}$SFR79 < 0), which agrees with our findings. It also indicates a recently enhanced SF on the periphery: that could be consistent with our results in some cases, for instance, if the SFR peaked between the past 800 and 5~Myr and remained higher at 5~Myr than at 800~Myr. This metric also does not identify the headlight cloud, which could have occurred if the SFR was much higher in that area $\sim$800~Myr ago, with an upturn between the past 5--200~Myr. Again, if there was a major boost in SF $\sim$1~Gyr ago as mentioned in \cite{MacArthur_09}, this scenario would be possible.

Generally, our \dsfr\,index partly agrees with the $\log_{10}$SFR79 diagnostic, although, the differences could be due to varying timescales. It is, however, difficult to draw any meaningful conclusions about $\log_{10}$SFR79 in NGC~628 due to a large number of missing pixels due to weak H$\delta$ absorption lines. The discrepancies between these SFR change metrics could be reconciled if there was a major increase in SF $\sim$1~Gyr across almost the entire disk, that would be comparable to the current SFR in the headlight cloud.

\section{Summary and Conclusions}\label{sect:conclusions}
Star formation (SF) activity in nearby galaxies varies strongly both between the galaxies and also on resolved scales inside one galaxy. These changes are important for our understanding of galaxy evolution and are imprinted in star formation histories (SFHs) of individual galaxies. One way to study the SFH observationally is to compare the changes in the star formation rate (SFR) occurring on different timescales.

In this paper, we studied SFR changes in a nearby spiral galaxy NGC~628. Using \texttt{CIGALE}, we defined a SFR change diagnostic as a ratio of the SFR averaged over the past 5 and 200~Myr, \dsfr, probed by the H$\alpha-$FUV colour. 
The main fin\-dings of this work are:
\begin{enumerate}
    \item Our \dsfr\ indicator shows that NGC~628 is overall going through a recent suppression of the SFR, albeit at a slower rate along the spiral arms. We also successfully managed to identify a strong burst that corresponds to the headlight cloud, that is, the large molecular gas cloud being destroyed by young massive stars \citep{headlight_cloud}. 
    
    \item \dsfr\,follows the available molecular gas reservoir, while the agreement between the \sfe\ and the molecular gas is not as clear. 
    In addition, \dsfr\, increases with $\Sigma$SFR at fixed $\Sigma$M(H$_2$), with the strongest SF enhancements occurring where the highest concentration of the SFR and molecular gas mass surface density is present, as shown in Figure \ref{fig:rKS}.

    \item Looking at the spiral arm structure,  \dsfr, again shows a better agreement than the \sfe. We compared the distributions of the two parameters in the arms and interarms using the Kolmogorov-Smirnov test and found that \dsfr\,is  sensitive to the spiral arm structure with high statistical significance (p-value = $1.67\times 10^{-15}$), unlike the \sfe (p-value = 0.25). From this we conclude that the spiral density wave concentrates gas in the spiral arms in NGC~628, leading to an increase in the recent SF, but does not enhance the \sfe, in line with previous findings \citep{foyle_10, kreckel_16}. 
    
    \item We also examined if an increase in pressure within the galactic disk, estimated by the hydrostatic pressure, can boost recent SF and \sfe. \dsfr\,appeared to react expectedly to such an increase resulting in elevated \dsfr. The \sfe, on the other hand, showed a decreasing trend with pressure. In NGC~628, the pressure grows towards the centre of the galaxy due to the concentration of gas and stars, indicating that the lower \sfe\ at the central, higher pressures may result from the stabilisation of the gas by the stellar bulge. 
    

    \item Lastly, we compare our findings to other tracers of recent SFR change: the $D\text{n}$(4000) break
    and $\log_{10}$SFR79 from \cite{wang_lilly_20}. There are some discrepancies between \dsfr\,and $\log_{10}$SFR79, as $\log_{10}$SFR79 predicts large areas that recently have experienced an increase in the SF. The 4000\AA\, break also indicates that most of the galaxy has recently experienced a significant SF increase, comparable to the current SFR in the headlight cloud. These discrepancies, however, could be attributed to the different timescales probed by these SFR probes.

In conclusion, we find \dsfr\,to be a suitable probe for recent SFH, as it correlates to the molecular gas presence, spiral arm structure, and disk pressure, and can therefore reflect recent SFR changes. We also point out that NGC~628 may not be the most optimal galaxy to study recent SFR changes, given that its SFR has mostly been going down in the recent past. For this reason, we would like to extend the galaxy sample in the future, to probe a wider range of galactic environments where more pronounced star formation activity changes may be happening at the present time. We intend to explore other galaxies in the PHANGS-MUSE survey and combine it with the IR observations required for the dust attenuation corrections. This will provide up to eleven additional star-forming spiral galaxies to test the \dsfr\, diagnostic on.
\end{enumerate}

\section*{Acknowledgements}
We thank the anonymous referee for their comments that significantly improved the quality of the paper. IDL acknowledges support from ERC starting grant \#851622~DustOrigin.This research has made use of the SIMBAD database, operated at CDS, Strasbourg, France. Packages used and not cited in the main text: seaborn \citep{seaborn}; montage is funded by the National Science Foundation under Grant Number ACI-1440620, and was previously funded by the National Aeronautics and Space Administration's Earth Science Technology Office, Computation Technologies Project, under Cooperative Agreement Number NCC5-626 between NASA and the California Institute of Technology.

\section*{Data Availability}
The data used in this article are publicly available through the ESO archive, NED, and DustPedia databases; any additional data will be shared upon reasonable request to the corresponding author.
 


\pagebreak
\bibliographystyle{mnras}
\bibliography{references} 


\bsp	
\label{lastpage}
\end{document}